\documentclass[12pt,preprint]{aastex}
\usepackage{mathrsfs}
\usepackage{appendix}
\usepackage{ulem}
\usepackage{natbib}
\def\gsim{\ \raise 3pt \hbox{$>$} \kern -8.5pt \raise -2pt \hbox{$\sim$}\ }
\def\lsim{\ \raise 3pt \hbox{$<$} \kern -8.5pt \raise -2pt \hbox{$\sim$}\ }
\def\gdf{\mathscr{G}}
\renewcommand{\emph}{\textit}

\begin{document}
\title{Fitting FFT--derived Spectra: Theory, Tool, and Application to Solar Radio Spike Decomposition}

\author{Gelu M. Nita\altaffilmark{1}, Gregory D. Fleishman\altaffilmark{1}, Dale E. Gary\altaffilmark{1}, William Marin\altaffilmark{2}, and Kristine Boone\altaffilmark{3}}

\altaffiltext{1}{Center For Solar-Terrestrial Research, New Jersey
Institute of Technology, Newark, NJ 07102, USA}
\altaffiltext{2}{Binghamton University, Vestal, NY 13850, USA}
\altaffiltext{3}{University of Calgary, Calgary, Canada}

\begin{abstract}

Spectra derived from fast Fourier transform (FFT) analysis of time-domain data intrinsically contain statistical fluctuations whose distribution depends on the number of accumulated spectra contributing to a measurement.
The tail of this distribution, which is essential for separation of the true signal from the statistical fluctuations, deviates noticeably from the normal distribution for a finite number of the accumulations. In this paper we develop a theory to properly account for the statistical fluctuations when fitting a model to a given accumulated spectrum. The method is implemented in software for the purpose of automatically fitting a large body of such FFT-derived spectra. We apply this tool to analyze a portion of a dense cluster of spikes recorded by our FST instrument \citep{FST} during a record-breaking event \citep{Cerruti} that occurred on 06 Dec 2006. The outcome of this analysis is briefly discussed.

\end{abstract}

\keywords{}

\section{Introduction}

Digitized time domain signals with a given instantaneous bandwidth are often analyzed at high spectral resolution using the Fast Fourier Transform (FFT) \citep{Benz_etal_2005b, Mannan_etal_2000, Heydt_etal_1999}. As we show below, fitting FFT-derived spectra with a model function can be challenging because of their non-Gaussian statistical fluctuations, which depend on both the true signal and any sources of added noise.  The exact properties of the fluctuations depend on the number of raw FFT spectra accumulated prior to analyzing the spectrum.

This study is particularly motivated by a unique observation of the record-breaking \citep{Cerruti} solar flare of 06 December 2006 by our FASR Subsystem Testbed \citep[FST--][]{FST} instrument, which revealed a rich variety of narrowband coherent emissions, including dense clusters of narrowband radio spikes. Solar radio spikes \citep{Droge_1977, Slottje_1978, Staehli_Magun_1986, Benz_1986} are a most intriguing type of solar coherent radio emission \citep{Slottje_1981, Guedel_Benz_1988, Isliker_Benz_1994, Benz_etal_2005}, and have attracted attention both observationally and theoretically because of their unique properties and potential diagnostic value \citep{Elgaroy_Sveen_1973, Benz_1985, Csi_Benz_1993, Fl_Meln_1998,spikes, Fl_2004, Dabrowski_etal_2005, Rozh_etal_2008, Benz_etal_2009, Dabrowski_etal_2011}. For example, the spectral bandwidth of spikes is typically 1\% or less of the frequency at which they occur \citep{Elgaroy_Sveen_1973, Benz_1986, Csi_Benz_1993, Messmer_Benz_2000, Rozh_etal_2008, Nita_etal_2008}, which implies that some controlling parameter of the spike source can in principle be determined with a corresponding precision of 1\% or better, which would be superior for remote diagnostics of the solar corona.

This diagnostic potential can be realized, however, only if the spike emission mechanism is solidly established and the corresponding source model is well understood. Although reports in the literature often favor the electron cyclotron maser \citep[ECM, e.g.,][]{Treumann_2006} emission mechanism \citep{Elgaroy_Sveen_1973, Stepanov_1978, Holman_etal_1980, Melrose_Dulk_1982, Aschwanden_1990, Gary_etal_1991, Guedel_Zlobec_1991, Fl_Meln_1998, Stupp_2000, spikes, Rozh_etal_2008}, it is not yet clear whether all observed spikes are due to a single mechanism, or perhaps have multiple causes \citep[e.g.,][]{Altyntsev_etal_2003, Meshalkina_etal_2004, Chernov_etal_2006, Magdalenic_etal_2006}. The latter is supported by some observations, e.g. by observations of post-flare spikes  \citep{Benz_etal_2002} originating from spatial locations lacking the relatively strong magnetic field required for ECM to operate. One way to test the hypothesis that different mechanisms are involved in different events is to examine the statistical properties of spikes in a uniform way from one event to another.

Irrespective of the coherent emission mechanism involved, the phenomenon of spike emission requires significant wave growth, driven by a kinetic instability in fast electrons. Various regimes of such instability result in different statistical distributions of spike parameters such as amplitude, lifetime, or bandwidth. Thus, getting a reliable distribution of these spike parameters from observations can shed light on both the production mechanism and on the fast electron distribution over energy and pitch-angle. In addition, spike parameter distributions contain information on global properties of the spike cluster source, such as its fragmentation into sources of individual spikes \citep{Benz_1985}, and the level of magnetic irregularities (turbulence) in the source \citep{Rozh_etal_2008, Nita_etal_2008}.

With the above motivations, we seek a uniform method to derive true spike parameters from spectral observations, which in the case of dense spike clusters requires reliable decomposition into individual spikes in the presence of statistical fluctuations that can masquerade as spikes. We emphasize that this is a nontrivial task given that the spike bandwidth can be comparable to the instrument spectral resolution, while the spike amplitude can be comparable to (or less than) the typical amplitude of statistical fluctuations.

Typically, an FFT-based instrument observes a sequence of $2n$ time-domain samples, which are converted to $n$-channel raw spectra and then accumulated over $M$ such spectra to get a final output spectrum. This averaging reduces the amplitude of statistical fluctuations, but at the same time degrades the temporal or spectral resolutions. Therefore, there is a trade-off in the number of channels $n$ and accumulations $M$ for a given application.  Fortunately, our FST observations offer a unique data set for studying the interplay between the spectral resolution and the level of statistical fluctuations, because the instrument directly records digitized time-domain samples obtained with 1~ns resolution.  These are organized in 100~$\mu$s contiguous blocks, which are separated by 19.9~ms time-domain gaps needed to read out and store the previously acquired contiguous block \citep{FST}. Direct recording of time-domain data allows one to post-process each contiguous time-domain block with different FFT settings (different values of $n$ and $M$), to obtain accumulated spectra with adjustable spectral resolution and accumulation length.

In this study we take advantage of this flexibility to identify the optimal spectral resolution needed to perform spike decomposition and obtain reliable statistical distributions of spike parameters. To achieve this goal we developed a theory describing properties of the statistical fluctuations for $M$-accumulate raw FFT spectra, tested it on simulated data, and created a software tool capable of automatically fitting the corresponding data. This tool has been employed in the analysis of FST data in order to get the amplitude and bandwidth distributions of the observed spikes, both of which turn out to be asymmetric distributions with prominent power-law tails toward high amplitudes or large bandwidths.
The open source code of this tool has been made publicly available as part of the SolarSoftWare (SSW-\emph{http://www.lmsal.com/solarsoft/}) scientific software repository, and is being located in the SSW distribution tree at \emph{..\textbackslash\textbackslash ssw \textbackslash radio\textbackslash ovsa\textbackslash idl\textbackslash fitting\textbackslash spike\_explorer.pro}.
In section~\ref{motivation} we describe the general problem in using FFT data to measure reliable parameters of the observed spikes.  The basic approach is to add spectral-peak model components representing individual spikes until an overall minimum $\chi^2$ criterion is met.  We find that a correct determination of $\chi^2$ requires a new theoretical treatment of the noise statistics for FFT-derived spectra, which we provide in section~\ref{FFTstat}.  To actually apply the theory to our problem requires not only estimating errors for a given model, but also determining whether and where to place additional model components.  A novel approach based on statistical likelihood of the run-length of regions of mismatch between data and model is developed in section~\ref{composite}.  In section~\ref{algorithm} we describe the model-fitting algorithm using simulated data.  Finally, the algorithm is applied to the 6 December 2006 event in section~\ref{data}, and the results are discussed in section~\ref{discussion}.

\section{Statement of the Problem}
\label{motivation}
\begin{figure}
\epsscale{0.4}\plotone{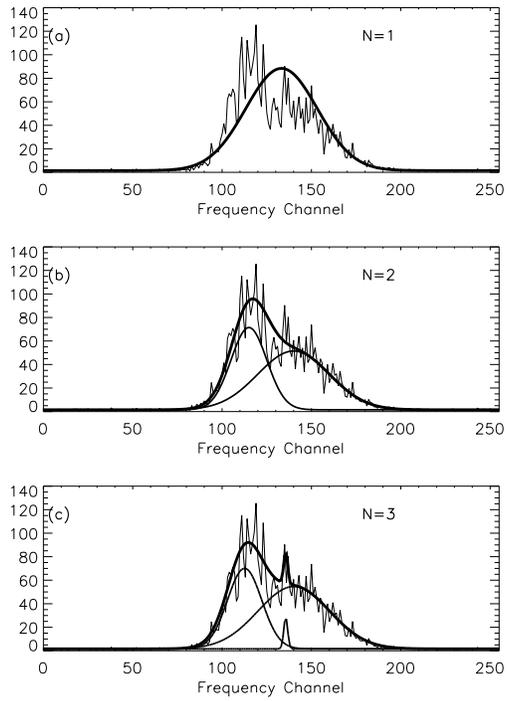}
\caption{\label{fitproblem} Least-squares solution (thick line) for the same FFT-derived spectrum based on three different spectral models assuming superposition (thin lines) of a flat background and one, two, and three Gaussian peaks (panels a, b, and c, respectively). }
\end{figure}

To illustrate the problem we face in this study, consider the simulated spike cluster shown in Figure~\ref{fitproblem}. The problem is to construct a model that adequately fits the data, consisting of one or more overlapping spectral components, each of which we take here as being adequately described by a Gaussian of adjustable amplitude and width.  The justification for this choice will be given in Section~\ref{algorithm}, but the general algorithm is independent of the specific form of the model.  The figure presents three standard least-squares solutions obtained from fitting an FFT-derived spectrum with different spectral models assuming superpositions of a flat background and one, two, or three Gaussian peaks, i.e.
\begin{equation}
s=\xi+\sum_{k=1}^N\alpha_k e^{-\frac{1}{2}\left(\frac{x-\beta_k}{\gamma_k}\right)^2},
\end{equation}
where $N=1, 2$ or $3$, and  $\{\xi, \alpha_k, \beta_k$, $\gamma_k\}$ (that is, background level, amplitude, position and width), are the unknown free parameters corresponding to each component.  It may appear that there are dozens of spikes in this cluster.  However, because the mean and noise are related for FFT-derived spectra, the noise is largest in channels where the intensity is greatest.  A statistically-based algorithm designed for the specific case of FFT-derived noise statistics is needed to assess whether a deviation of the model from the data can be explained by noise, or instead requires placement of an additional component.

The next two sections use the specialized language and notation of statistics theory to formally derive the relevant relationships necessary to address the problem outlined above.  Readers interested only in the application and results can skip to section~\ref{data}.

\section{Statistical Properties of FFT-derived Spectra}
\label{FFTstat}
\subsection{Statistical Fluctuations}

Statistics of the FFT-derived spectra can be formulated in terms of the standard \emph{Gamma} probability distribution function (PDF)
\begin{eqnarray}
\label{gamma}
\gdf(x;k,\theta)=\frac{x^{k-1}e^{-\frac{x}{\theta}}}{\theta^k\Gamma(k)},
\end{eqnarray}
where $\Gamma(z)=\int_0^\infty{t}^{z-1}e^{-t}dt$ is the Euler's Gamma function, and $k$ and $\theta$ are, respectively, the shape and scale parameters of the distribution, which determine its mean, $\mu=k\theta$, variance, $\sigma^2=k\theta^2$, skewness, $\alpha_3=2/\sqrt{k}$, and kurtosis $\beta_2=3+6/k$.

In particular, $\gdf(y_j,1,\mu_j)$ represents an exponential PDF of mean $\mu_j$ and variance $\mu_j^2$, which defines the statistical properties of a single (\emph{j}--indexed) channel of a raw FFT-derived spectrum \citep{RFI1,RFI2}, while $\gdf(S_j,M,\mu_j)$ represents the PDF corresponding to the sum of $M$ such random samples, which defines the statistical properties of each channel in an accumulation of $M$, \emph{i}--indexed, consecutive raw FFT-derived spectra \citep{RFI2},

\begin{equation}
\label{S1}
S_j=\sum_{i=1}^M y_j^{(i)}.
\end{equation}

Hence, the frequency--dependent population means and variances of such an accumulation are given by $s_j=M\mu_j$ and $\sigma_j^2=s_j^2/M$, respectively.
The channel--dependent parent distributions of the accumulated power $S_j$ may be expressed in terms of their individual statistical means $s_j$ and their common accumulation number $M$ as
\begin{equation}
\label{accpdf}
\gdf(S_j;M,s_j/M)=\frac{M^M}{\Gamma(M)}\frac{1}{s_j}\rho_j^{M-1}e^{-M\rho_j},
\end{equation}
where we introduce the sample-to-mean ratio $\rho_j=S_j/s_j$sout{ to denote the ratios between the random variables $S_j$ and their corresponding population means $s_j$}, which we are going to eventually derive from the data in the form of
the Sample to Model Ratio (SMR) estimator,
\begin{equation}
\label{SMR}
\widehat{\rho_j}=\frac{S_j}{\widehat{s_j}},
\end{equation}
with $\widehat{s_j}$ being the most likely solution for $s_j$.

\subsection{Maximum Likelihood Estimate of Spectral Shape Model Parameters}

If a model function $\widehat{s_j}$, described by a set of yet to be determined free parameters $\{p_k\},(k=\overline{1,\nu})$, is a true solution for the parent population means $s_j$ corresponding to the FFT-derived spectral points $S_j,(j=\overline{1,N})$,  the PDF given by equation~(\ref{accpdf}) can be used to build the associated likelihood function \citep{Bev}
\begin{equation}
\label{likelihood}
\mathscr{L}(p_1,p_2,...,p_\nu)=\prod_{j=1}^N\gdf[S_j;M,\frac{1}{M}\widehat{s_j}(p_1,p_2,...,p_\nu)],
\end{equation}
which represents the conditional probability density function associated with the observation.

The maximum likelihood estimates of the model parameters $\{p_k\}$ are obtained by maximizing the likelihood function or, equivalently, by minimizing the more mathematically convenient negative log--likelihood function \citep{Bev}
\begin{eqnarray}
\label{ll}
&&\lambda(p_1,p_2,...,p_\nu)\equiv-2\ln\left[\mathscr{L}\right]=\\\nonumber
&&2M\sum_{j=1}^N\left\{\frac{S_j}{\widehat{s_j}}+\ln(\widehat{s_j})
-\left(1-\frac{1}{M}\right)\ln\left(S_j\right)+c(M)\right\},
\end{eqnarray}
where $c(M)=\ln\left[\Gamma(M)\right]/M+\ln(M)$ is a channel--independent constant.
For Gaussian statistics, $\lambda$ is equivalent to the standard $\chi^2$  function
\begin{equation}
\label{chisqr}
\chi^2=\sum_{j=1}^Nw_j\left(S_j-\widehat{s_j}\right)^2,
\end{equation}
where  $w_j$ are the statistical weights, related by $\chi^2=\lambda+\rm{constant}$.

\subsection{Goodness-of-Fit Assessment}

A major drawback of maximum likelihood is that it does not provide a standard method for assessing goodness of fit. The standard $\chi^2$-based goodness-of-fit assessment may not be valid unless the $\chi^2$ parameter associated with the problem follows a $\chi^2$ PDF with $\nu$ degrees of freedom, i.e.
\begin{equation}
\label{chi2pdf}
\texttt{PDF}[\chi^2(\nu)]=\gdf\left(\chi^2;\frac{\nu}{2},2\right),
\end{equation}
where $\nu=N-n$ is given by the difference between the number of the data samples $N$ and the number of free parameters $n$ of the model function.

The expectation of the random variable $\chi^2$ is $E(\chi^2)=\nu$ or, equivalently, $E(\chi_\nu^2)=1$, where $\chi_\nu^2\equiv\chi^2/\nu$ is the reduced chi-square associated with the re-normalized distribution
\begin{equation}
\label{normchi2pdf}
\texttt{PDF}[\chi_\nu^2]=\gdf\left(\chi_\nu^2;\frac{\nu}{2},\frac{2}{\nu}\right).
\end{equation}

To investigate whether the standard $\chi^2$ function (\ref{chisqr}) works for FFT-derived spectra,
we neglect all sources of noise other than the FFT statistical fluctuations, so that statistical weights $w_i$ in equation~(\ref{chisqr}) are described by the variance of the parent \textit{Gamma} distribution of the random variables $s_j$, i.e. $w_j=1/\sigma_j^2=M/s_j^2$, which depend exactly on the quantities to be estimated.  We estimate these statistical weights from the model itself as
\begin{equation}
\label{model_w}
w_j=\frac{M}{\widehat{s_j}^2}.
\end{equation}
Substituting  these yet unknown weights to equation~(\ref{chisqr}) we find
\begin{equation}
\label{chisqr1}
\chi^{2(I)}\equiv M\sum_{j=1}^N\left(1-\widehat{\rho_j}\right)^2,
\end{equation}
with $\widehat{\rho_j}$ defined by equation~(\ref{SMR}).  The superscript ($I$) is used to distinguish this expression for $\chi^2$ with another that we will develop shortly.

Clearly, the solution provided by the minimization of equation~(\ref{normchi2pdf}) is located in the vicinity of the solution provided by least-squared minimization of equation~(\ref{chisqr1}). Indeed, both minimization problems seek a spectral model estimate $\widehat{s_j}$ that globally minimizes the deviation of $\widehat{\rho_j}$ from unity.  The SMR estimator $\widehat{\rho_j}$ is a random variable belonging to the same parent population as $\rho_j$. The PDF of $\rho_j$ may be straightforwardly derived from equation~(\ref{accpdf}) through a simple scale transformation, leading to
\begin{equation}
\label{normpdf}
\gdf\left(\rho_j,M,\frac{1}{M}\right)=\frac{M^Me^{-M\rho_j}\rho_j^{M-1}}{\Gamma(M)},
\end{equation}
with mean 1 and variance $1/M$, cf. equation~(\ref{gamma}).

Making use of parent PDF (\ref{normpdf}) of $\rho_j$, the expectation of $\chi^{2(I)}$ (\ref{chisqr1}) can be computed under the assumption that the model function $\widehat{s_j}$ represents the true spectral shape $s_j$:
\begin{eqnarray}
\label{chi2I_expectation}
E\left[\chi^{2(I)}\right]&=&N.\\\nonumber
\end{eqnarray}

Using equation~(\ref{chi2I_expectation}), we may define the normalized functional form with unity expectation
\begin{eqnarray}
\label{chi2Inorm}
&&\chi^{2(I)}_N\equiv M\frac{1}{N}\sum_{j=1}^N\left(1-\widehat{\rho_j}\right)^2,\\\nonumber
\end{eqnarray}
or in the case of $n$ model parameters to be determined from the data, we replace $N$ with the number of
degrees of freedom $\nu=N-n$, as in the standard reduced-chi-square estimator \citep{Bev}, to arrive at our
final goodness-of-fit estimator
\begin{equation}
\label{redchisqr}
\chi_\nu^2\equiv M\frac{1}{\nu}\sum_{j=1}^N\left(1-\widehat{\rho_j}\right)^2.
\end{equation}

As shown in Appendix \ref{APP_CHI2PDF}, the PDF of $\chi_\nu^2$ may be approximated by the analytical expression
\begin{equation}
\label{chisqPDFapprox}
\texttt{PDF}\left[\chi_\nu^2\right]\approx\mathscr{G}\left[\chi_\nu^2;\frac{1}{\left(1+\frac{3}{M}\right)}\frac{\nu}{2},\left(1+\frac{3}{M}\right)\frac{2}{\nu}\right],
\end{equation}
which, in the limit of large accumulation length $M$, reduces to a classic chi-squared distribution normalized by its degrees of freedom, equation~(\ref{normchi2pdf}).
Using this approximation, the probability to observe a given $\chi_\nu^2$ or larger may be computed as
\begin{equation}
\label{pvalue}
P(\chi_\nu^2)=\frac{\gamma\left(\frac{M}{M+3}\frac{\nu}{2},\frac{M}{M+3}\frac{\nu}{2}\chi_\nu^2\right)}{\Gamma\left(\frac{M}{M+3}\frac{\nu}{2}\right)}.
\end{equation}
where
\begin{equation}
\label{igamma}
\gamma(a,z)=\int_z^\infty t^{a-1}e^{-t}dt
\end{equation}
is the incomplete \emph{Gamma} function.

\subsection{Validating Simulations}
\label{S_Sim}

\begin{figure}
\epsscale{0.6}\plotone{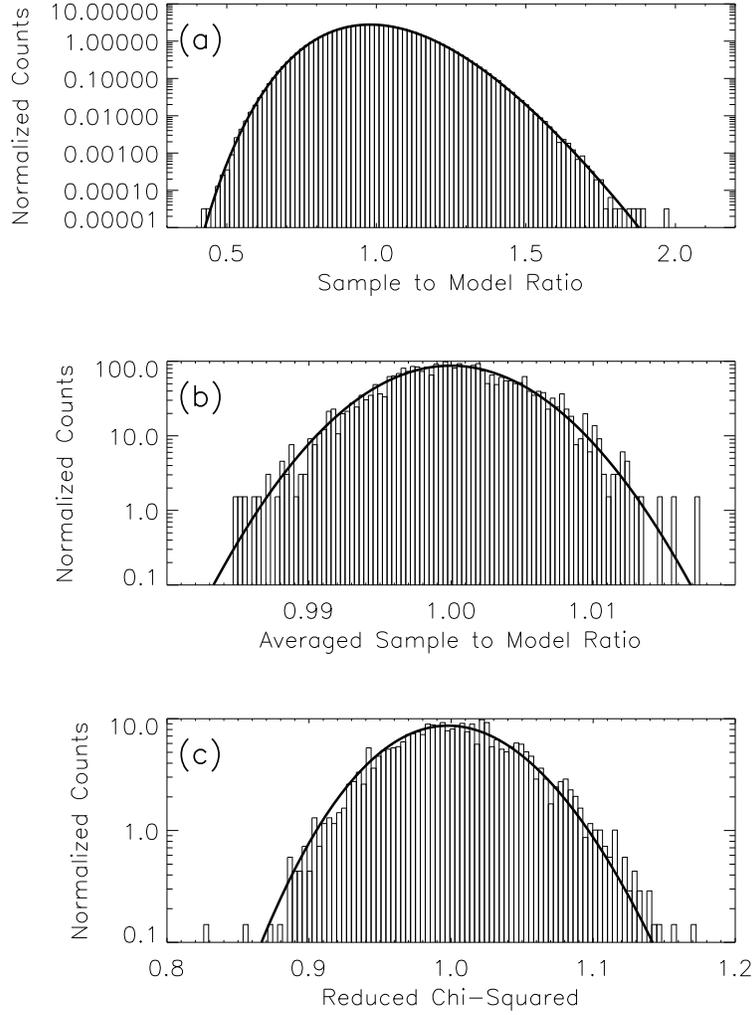}
\caption{\label{gofsim} Monte Carlo simulation of a set of 2 million $M=48$--$\rho_j$ random deviates divided into $2,000$ $N=1,000$--contiguous blocks. a) Observed $\rho_j$ density distribution (histogram) and its theoretical PDF (solid-thick line) given by equation~(\ref{normpdf}). b) Observed density distribution of a sequence of $2,000$ $\eta_i=(1/N)\sum_{j=1}^N\rho_j$ mean SMR values and its theoretical PDF (solid-thick line), $\gdf(\eta,MN,1/MN)$. c) Observed density distribution of $2,000$ $\chi_\nu^2$ values computed for each contiguous bloc and its theoretical PDF (solid-thick line) given by equation ~(\ref{chisqPDFapprox}).}
\end{figure}

In order to validate the theoretical derivations presented up to this point, and also for further reference, we have performed a Monte Carlo simulation in which we generated a sequence of 2 million $\rho_j$ random variables distributed according to a $\mathscr{G}(\rho_j,M,1/M)$ PDF, which corresponds to a structureless flat spectrum of unity mean. A particular value of $M=48$ has been chosen for illustration. Figure~\ref{gofsim}$(a)$ displays the density distribution of the numerically generated SMR data set, as well as a curve representing the corresponding theoretical PDF given by equation~(\ref{normpdf}).

The 2 million values have been divided into $2000$ contiguous blocks of length $N=1000$, from which are computed a sequence of $2000$ SMR means, $\eta_i=\sum_{j=1}^N\rho_j$. Figure~\ref{gofsim}$(b)$ displays the density distribution of the SMR means, as well as the corresponding theoretical PDF curve $\gdf(\eta,MN,1/MN)$.

Finally, equation~(\ref{redchisqr}) has been used to compute a set of $2000$ $\chi_\nu^2$ deviates, where $\nu\equiv N$, since there are no free parameters. Figure \ref{gofsim}$(c)$ displays the density distribution of the computed $\chi_\nu^2$ values, as well as the corresponding PDF curve of equation~(\ref{chisqPDFapprox}).

Visual inspection of Figure \ref{gofsim} shows good agreement between the numerically generated distributions and their theoretical expectations, not only in the case of the exact $\rho_j$ and $\eta_i$ PDFs, but also in the case of the $\chi_\nu^2$ PDF approximation.

\subsection{Least Squares Initial Guess of Spectral Shape Model Parameters}

Another drawback of the maximum likelihood method is its tendency to converge toward a local extreme, rather than an absolute one. Instead,
we use a least-squares minimization for the purpose of making an initial guess of the unknown parameters. However, the weighting scheme adopted in equation~\ref{model_w} is no good for this purpose, since the weights come from the model itself, which we do not initially know. To comply with the standard implementation of the least-squared minimization algorithm, which employs weighting coefficients $w_i$  derived from measurement errors, we employ the unbiased variance estimator $\widehat{\sigma_j^2}=S_j^2/(M+1)$ derived in Appendix \ref{sumvar} to assign the sample-based weights $w_j\rightarrow (M+1)/S_j^2$ based on the measured data samples. This is equivalent to minimizing the quantity
\begin{equation}
\label{chisqr2}
\chi^{2(II)}\equiv{(M+1)}\sum_{j=1}^N\left(1-\frac{1}{\widehat{\rho_j}}\right)^2,
\end{equation}
which pursues in a reciprocal space the same goal as the model-based weighing scheme, i.e. finding the model that globally minimizes the deviations of the same local SMR parameters from unity. However, the uncertainties associated with the estimated parameters provided by the local curvature at the point of absolute $\chi^2$ minimum \citep{Bev},
\begin{equation}
\label{chi2sigma}
\sigma_{p_k}^2=2\left[\frac{\partial^2\chi^2}{\partial{p_k^2}}\right]^{-1}
\end{equation}
are different for the two weighting schemes, which affects the most likely solution obtained, as quantitatively illustrated in Appendix \ref{comparison}.

\begin{figure}
\epsscale{1}\plotone{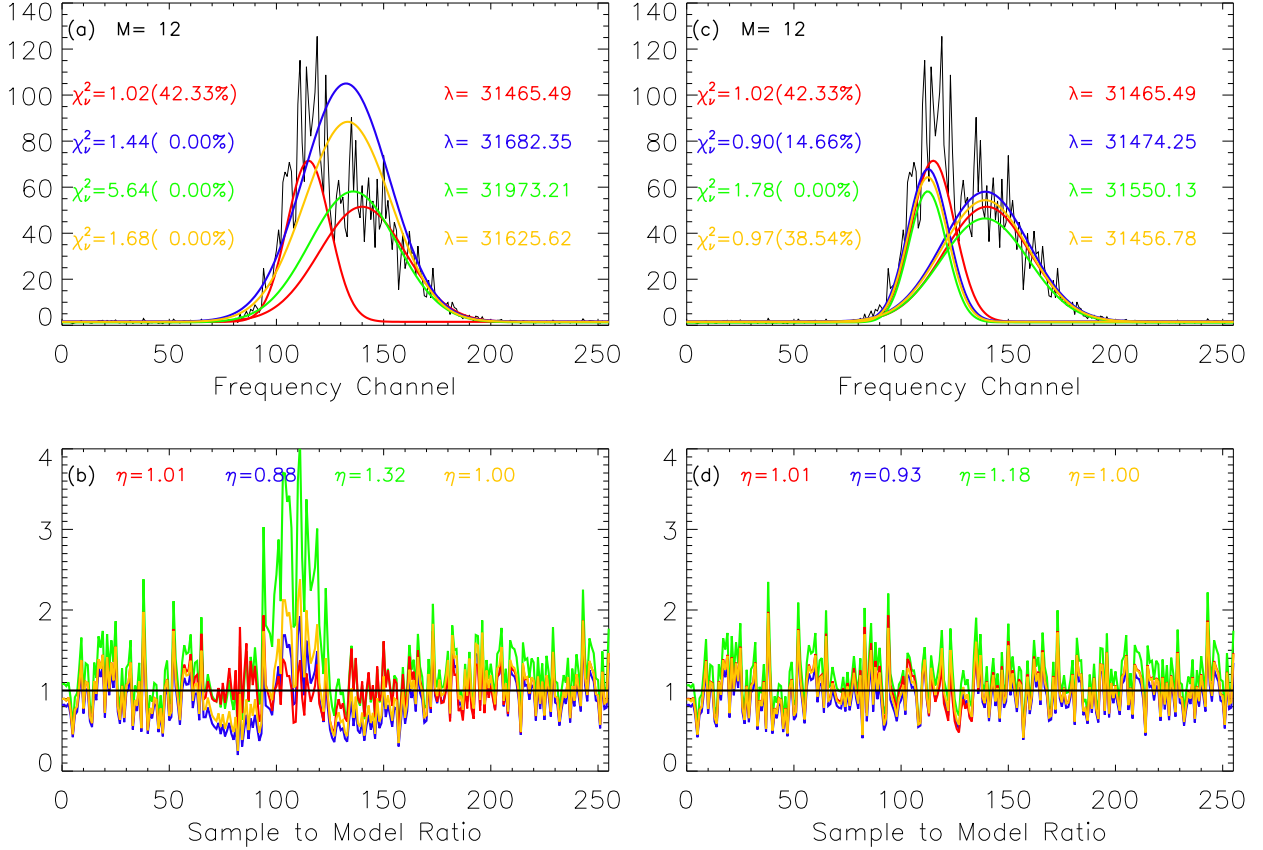}
\caption{\label{twospikes} Fitting results of a composite spectrum formed by superposition of two partially overlapped peaks under the assumptions of a one peak model (panels a and b) and two peak model (panels c and d). The exact (red),  $\chi^{2(I)}$ (blue), $\chi^{2(II)}$ (green) and $\lambda$ (yellow) solutions are shown in the upper panels, and their corresponding SMR deviations in the lower panels. The associated $\chi_\nu^2$, $\lambda$ and $\eta$ parameters, as well as the probabilities to have observed greater $\chi_\nu^2$ under the assumption of a correct spectral model are also displayed in corresponding colors. }
\end{figure}

Minimization of estimator equation~(\ref{chisqr2}) is the more practical of the alternatives.
Consider  the composite spectrum  illustrated in Figure \ref{twospikes}, which consists of  superposition of two partially overlapping Gaussian peaks, same as in Figure~\ref{fitproblem}, shown by the red curves. In order to fit the simulated, noise-contaminated spectrum (thin black line), calculated for an accumulation length $M=12$, one must first estimate the number of Gaussian peaks from the data. Figure \ref{twospikes}(a) illustrates the fitting results with an initial guess that consists of only a single peak. The $\chi^{2(I)}$ (blue), $\chi^{2(II)}$ (green) and $\lambda$ (yellow) minimizations lead to different solutions characterized by different goodness of fit parameters $\chi_\nu^2$ and different negative log--likelihoods $\lambda$. The $\lambda$ and $\chi^{2(I)}$ minimizations lead to solutions that have goodness-of-fit and likelihoods comparable with the true model function, while the $\chi^{2(II)}$ minimization leads to a solution with poorer goodness-of-fit and likelihood. However, it has the advantage of being closer to the true shape of one of the overlapping peaks, unlike the alternative solutions that provide estimates that are closer to the envelope of the overlapping peaks\footnote{
This reflects typical performance of each of these estimation methods: the $\chi^{2(II)}$ functional form gives, by design, more weight to the low amplitude data points, which are less affected by FFT noise than the peaks, while the $\chi^{2(I)}$  and $\lambda$ functional forms favor global solutions that are more consistent with the entire data set under the assumption that the true spectral model contains only one spectral peak.}. The same conclusion is supported by Figure \ref{twospikes}(b), which displays the local deviations $\rho_j$ and their averages over the entire spectrum $\eta$.
Now, the poorer $\chi^{2(II)}$ minimization signals the inadequacy of the model while the non-uniform pattern of the SMR deviations indicates the most appropriate location to make a model adjustment. By adding a new spectral peak to the model close to the suggested location, the results displayed in Figures \ref{twospikes}(c) and (d) are obtained. These panels reveal that in the case of a good model fit, $\chi^{2(I)}$  and $\lambda$ minimizations provide more accurate estimates of the true spectral peaks than the $\chi^{2(II)}$  minimization, as indicated by their $\chi_\nu^2$ and $\eta$ estimators being closer to unity. This suggests a strategy where values from $\chi^{2(II)}$  minimization are used to identify where new model components are needed, and $\chi^{2(I)}$ and $\lambda$ minimizations are used to assess the final model parameters once the model components are determined.  In practice, we find that $\lambda$ minimization yields the best results.

\section{Fitting Composite Spectra: Estimation of the Most Likely Spectral Model}
\label{composite}

As has been shown, $\chi^{2(II)}$-minimization is the most sensitive to a local data--model mismatch.  The presence of local SMR deviations greater than statistically expected can suggest the spectral location at which the spectral model should be amended (by adding one more peak, for example).
This offers a starting point for designing an adaptive, self-consistent spectral-fitting algorithm that should be capable of adding complexity to an initially crude spectral model until the remaining SMR fluctuations around unity are solely due to statistical fluctuations, within an acceptable confidence interval.

\subsection{Statistics of SMR Systematic Deviations}

The probability to observe a local SMR exceeding an observed value $\rho_0$  due to statistical fluctuations can be computed from equation~(\ref{normpdf}) to yield
\begin{equation}
\label{normprob}
p(\widehat{\rho_j}>\rho_0)=\frac{\gamma(M,M\rho_0)}{\Gamma(M)}.
\end{equation}
Consequently, the probabilities to observe a local SMR deviation above or below the mean, $p_a\equiv p(\widehat{\rho_j}>1)$ and $p_b\equiv p(\widehat{\rho_j}<1)$, respectively, are given by 
\begin{eqnarray}
\label{pab}
p_{a}&=&\frac{\gamma(M,M)}{\Gamma(M)}\\\nonumber
p_{b}&=&1-\frac{\gamma(M,M)}{\Gamma(M)}
\end{eqnarray}

To identify a strong deviation from the model, equation~(\ref{normprob}) alone may be sufficient.  In a more general case, however, especially for a smaller-amplitude but spectrally-resolved signal, we expect that the data will deviate from an imperfect model in some compact spectral region that will result in systematic deviations of the SMR estimator either above or below unity.

To quantitatively address this aspect of the fitting problem, we derive the PDF describing the probability of observing a compact region of a given length whose SMR deviates from unity solely as the result of statistical fluctuations. This will allow a decision on where and when a local improvement to the model function is needed, based on some desired probability of false alarms.

We first derive the discrete probability mass functions (PMF) for observing a SMR compact region of a certain size $n$ that is located above or below unity. The process of such compact regions of a given size occurring randomly can be described as a memoryless Bernoulli process in which any individual SMR in a random sequence may be above or below unity with the unequal probabilities $p_a$ and, respectively $p_b$ (equation~\ref{pab}).  The two size distributions should obey complementary geometrical distributions given by
\begin{eqnarray}
\label{nab}
f(n_a)&=&\left[1-\frac{\Gamma(M)}{\gamma(M,M)}\right]\left[\frac{\Gamma(M)}{\gamma(M,M)}\right]^{n_a}\\\nonumber
f(n_b)&=&\frac{\Gamma(M)}{\gamma(M,M)}\left[1-\frac{\Gamma(M)}{\gamma(M,M)}\right]^{n_b},
\end{eqnarray}
which represent the conditional probabilities of observing a run of SMR deviations, all on the same side of unity, of length $n_{a,b}$, terminated by any SMR deviation of the opposite sense.  We have confirmed this expectation with simulated FFT-derived spectral data.

In addition to the run-length criterion, we can examine the probability for an identified region to deviate in amplitude by more than statistically expected from the population mean of $\widehat{\rho_j}$ which we define as the mean of the $n$ contiguous individual SMR deviations that are all above, or all below, unity,

 \begin{eqnarray}
 \label{sab}
 \eta_a=\frac{1}{n}\sum_{j=j_0}^{j_0+n-1}[\widehat{\rho_j}\ge1],\\\nonumber
 \eta_b=\frac{1}{n}\sum_{j=j_0}^{j_0+n-1}[\widehat{\rho_j}\le1].
 \end{eqnarray}

To derive the PDFs of such means, we first obtain the conditional PDFs (CPDF) associated with each type of compact deviate by splitting the unconditional PDF defined by equation~(\ref{normpdf}) into two truncated \emph{Gamma} distributions, above and below unity, and normalizing them by the factors provided by equation~(\ref{pab}). Hence,
 \begin{eqnarray}
  \label{cnormpdf}
  \mathrm{CPDF}(\rho_j>1)&=&\frac{\Gamma(M)H(\rho_j-1)}{\gamma(M,M)}M^Me^{-M\rho_j}\rho_j^{M-1}\\\nonumber
  \mathrm{CPDF}(\rho_j<1)&=&\frac{\Gamma(M)H(1-\rho_j)}{\Gamma(M)-\gamma(M,M)}M^Me^{-M\rho_j}\rho_j^{M-1},
 \end{eqnarray}
where $H(x)$ represents the Heaviside unit step function.

Although the desired PDF cannot be obtained in closed form, we can nevertheless compute its moments in a standard way, which we do in Appendix \ref{appendix_moments}.  Equations~(\ref{cma}) and (\ref{cmb}) give the means and central moments $\mu_i$ of the random variables $\eta_a$ and $\eta_b$, respectively. These central moments may be used to compute, for a given set of parameters $M$ and $n$, the more commonly used standard moments, i.e. standard deviation $\sigma=\sqrt{\mu_2}$, the skewness $\alpha_3=\mu_3/\mu_2^{3/2}$, and  kurtosis $\beta_2=\mu_4/\mu_2^2$. For convenience, we will also use in the subsequent derivations the alternative standard parameter $\beta_1=\alpha_3^2=\mu_3^2/\mu_2^3$.

\subsection{Pearson Type I Approximations of the Mean SMR Distributions}
\label{P1}

Having determined the exact first four standard moments of the Mean SMR probability distribution functions, it may be shown (see Appendix \ref{appendix_P1vMC} for a detailed quantitative analysis) that the Pearson Type I PDF \citep{Pearson} is a generally valid approximation, at least for large run-lengths $n\gg1$, for the true PDF describing the parent populations of the $\eta_a$ and $\eta_b$ random variables.

\begin{figure}
\epsscale{0.8}\plotone{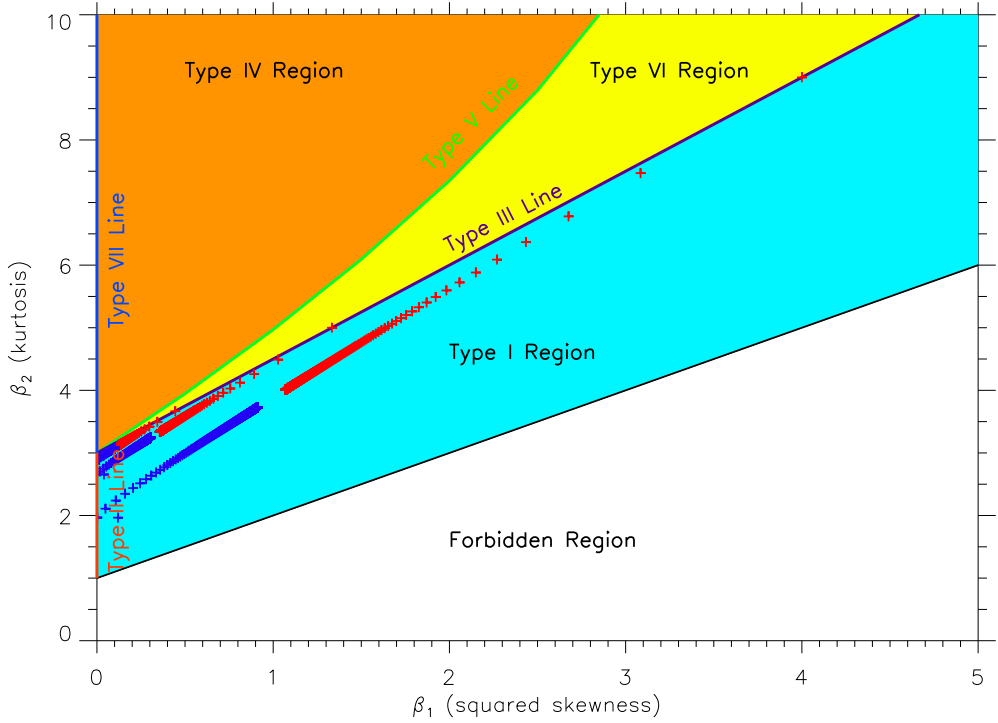}
\caption{\label{pearsonregions} Kurtosis $(\beta_2)$ and squared skewness ($\beta_1$) as computed from Eqn. \ref{cma} (red symbols) and Eqn. \ref{cmb} (blue symbols) for $n=1,3,$ and $9$, and $M$ ranging from $1$ to $1000$. The symbols are overlaid on a color-coded diagram which indicates the parameter regions corresponding to each of the $7$ probability distribution types that form the Pearson system. All Pearson lines and regions converge in the $\{\beta_1=0,\beta_2=3\}$ point, which corresponds to the normal distribution. {As $M$ increases, the red and blue symbols corresponding to the same value of $n$ converge from both sides of the same line toward a central point $\{\beta_1,\beta_2\}$. As $n$ increases, the central point corresponding to each such line approaches the point of the diagram defining the normal distribution.}}
\end{figure}

This is suggested by Figure \ref{pearsonregions}, which displays the relationship between the kurtosis $(\beta_2)$ and squared skewness ($\beta_1$) as computed from equation~(\ref{cma}) (red $+$ symbols) and equation~(\ref{cmb}) (blue $+$ symbols) for  $n=1,3,$ and $9$, and $M$ ranging from $1$ to $1000$. The symbols are overlaid on a color-coded diagram that indicates the parameter regions corresponding to each of the $7$ probability distribution types that form the Pearson system. The Pearson Type I region is upper bounded by the line, $2 \beta_2 - 3 \beta_1 - 6 =0$, and lower bounded the line $\beta_2 -\beta_1 -1=0$. The $n$ have been chosen to illustrate the influence of the two free parameters $n$ and $M$ on the shape of the $\eta_a$ and $\eta_b$ distributions. The three parallel symbol stripes, which correspond to different values of $n$, indicate that the skewness range decreases as $n$ increases, while, for a fixed $n$, the $\eta_a$ and $\eta_b$ distributions converge toward the same skewness as $M$ increases. However, unless the length of a compact SMR region is fairly large, $n\gg1$, its parent PDF cannot be accurately approximated by a purely Gaussian distribution (which corresponds to the point on the $y$-axis $\{\beta_1=0,\beta_2=3\}$).

In its most general form, the Pearson Type I PDF is a generalized \emph{Beta} distribution having the functional form
\begin{eqnarray}
\label{betapdf}
f(x,a,b,p,q)=\frac{\Gamma(p + q)(x - a)^{p - 1} (b - x)^{q - 1}}{\Gamma(p) \Gamma(q)(b - a)^{p + q - 1}};\\\nonumber
a\le x\le b,
\end{eqnarray}
where $a$ and $b$ define the endpoints of the limited domain of definition, and $p$ and $q$ are two parameters that determine the shape of the distribution.
These four free parameters have to be determined such that the first four moments of the true distributions are matched by the first four moments of their respective Pearson Type I approximations.

The solution to this problem is given by
\citep{beta},
\begin{eqnarray}
\label{betaparms}
a &=& \mu - \frac{\sigma}{2} \frac{p \sqrt{d}}{p + q}\nonumber\\
b &=& \mu + \frac{\sigma}{2} \frac{q \sqrt{d}}{p + q}\nonumber\\
p &=& \frac{r}{2} \left(1 - (r + 2) \sqrt{\frac{\beta_1}{d}}\right)\\\nonumber
q &=& \frac{r}{2} \left(1 + (r + 2) \sqrt{\frac{\beta_1}{d}}\right)\\\nonumber
r &=& \frac{6 \beta_2 - \beta_1 - 1}{6 + 3 \beta_1 - 2 \beta_2}\\\nonumber
d &=& (r + 2)^2 \beta_1 + 16 (r + 1)\nonumber,
\end{eqnarray}
where all quantities are expressed in terms of the known standard moments $\mu,\sigma,\beta_1$ and $\beta_2$.

The probability for the mean of a compact SMR region, located above or below unity, to lie above or, respectively, below a certain threshold, $t$, may be expressed in terms of the cumulative probability function of the \emph{Beta} distribution, $p(t)=\int_t^\infty f(x,a,b,p,q)dx$, which may be conveniently written as
\begin{eqnarray}
\label{betaprob}
p(t)=\left\{\begin{array}{ll}
       1 & t\le a \\
       I\left(\frac{1}{2} (1 + \frac{t_0}{\sigma t_1}), 1 + t_3 - \frac{t_4}{t_1}, 1 + t_3 + \frac{t_4}{t_1}\right) & a<t<b\\\nonumber
       0 & t\ge b,
       \end{array}\right.\\
\end{eqnarray}
where
\begin{equation}
\label{ibeta}
I(z,\alpha,\beta)=\frac{\int_0^z\xi^{\alpha-1}(1-\xi)^{\beta-1}d\xi}{\int_0^1\xi^{\alpha-1}(1-\xi)^{\beta-1}d\xi}
\end{equation}
is the regularized incomplete \emph{Beta} function, and
\begin{eqnarray}
\label{tparms}
t_0&=&(4\beta_2-6\alpha_3^2-12)(t-\mu) + \sigma\alpha_3( \beta_2+3)\\\nonumber
t_1&=&\sqrt{\alpha_3^2(\beta_2^2+78\beta_2-63) - 32(\beta_2-3)\beta_2-36\alpha_3^4 }\\\nonumber
t_3&=&\frac{5\beta_2- 6\alpha_3^2-9}{3\alpha_3^2 - 2\beta_2+6}\\\nonumber
t_4&=&\frac{3\alpha_3(\alpha_3^2 - \beta_2+1)(\beta_2+3)}{3\alpha_3^2 - 2\beta_2+6},
\end{eqnarray}
where the skewness $\alpha_3$  and not $\sqrt{\beta_1}$ has been explicitly used for properly taking its sign into consideration.

To determine whether the deviation of an observed SMR mean is statistically significant, one may use equation~(\ref{betaprob}) for deviations above unity, or its complement $1-p(t)$, for deviations below unity. Exceeding an acceptable probability of false alarm (PFA), such as the standard $3\sigma$ PFA of $0.13499\%$ for the normal distribution, would signal a statistically significant deviation. However, unlike the normal distribution, the Pearson Type I distribution has a limited domain outside of which the PFA is identically 0. In the following, unless otherwise stated, we will use this most-conservative 0\% PFA by setting the thresholds to the minimum and maximum theoretically allowed deviations, i.e $\eta_b = a$ and $\eta_a = b$, respectively.

In conclusion, we can combine both conditions as a multivariate probability function, the product of the probability to have a region with a given length $n$ from equation~(\ref{nab}) and the probability of exceeding threshold $t$ from equation~(\ref{betaprob}), i.e.
\begin{eqnarray}
\label{absolute}
p(t,n)&=&\left\{\begin{array}{ll}
       0&t\le a<1\\\\
       \left[\frac{\Gamma(M)}{\gamma(M,M)}\right]\left[1-\frac{\Gamma(M)}{\gamma(M,M)}\right]^{n}
       \left[1-I\left(\frac{1}{2} (1 + \frac{t_0}{\sigma t_1}), 1 + t_3 - \frac{t_4}{t_1}, 1 + t_3 + \frac{t_4}{t_1}\right)\right] & a<t<1 \\\\
       \left[1-\frac{\Gamma(M)}{\gamma(M,M)}\right]\left[\frac{\Gamma(M)}{\gamma(M,M)}\right]^{n}
       \left[I\left(\frac{1}{2} (1 + \frac{t_0}{\sigma t_1}), 1 + t_3 - \frac{t_4}{t_1}, 1 + t_3 + \frac{t_4}{t_1}\right)\right] & 1<t<b\\\\
       0 & t\ge b,
       \end{array}\right.
\end{eqnarray}
where the parameters $ a,b,t_0,t_1,t_3,t_4$ and $\sigma$ depend only on $M$, $n$, and the position of $t$ relative to unity.  This expression provides accurate tail probabilities associated with the SMR mean $t$ over a compact region of length $n$ and therefore may be used in a fitting algorithm to decide whether to add another component to the fit model. We note that this is a general result that does not depend in any way on the choice of the fit components. In what follows, we will decompose the spectrum into Gaussian components, but one could as easily choose wavelets, lorentzians, Voigt,  or other forms.

\section{An algorithm for fitting spectral data with Gaussian components}
\label{algorithm}

This section describes a self-adaptive algorithm specifically tailored for the problem stated in section~\ref{motivation}---that is, estimating the most probable superposition of Gaussians that fit an observed spectrum obtained by accumulating a known number of raw FFT-derived spectra. {Our model choice (superposition of positive-amplitude Gaussians) necessarily means that the algorithm described below flags only positive SMR deviations to indicate where to add components.  The theory above, however, does not make any assumption about the form of the model, and for some models (e.g. superposition of wavelets, or spectra where both absorption and emission are expected) it could be appropriate to flag both positive and negative SMR deviations for adding components.} The selection of {positive} Gaussian spectral components is not arbitrary {for the case of radio spike bursts}. Indeed, as has been said in the Introduction, a coherent emission mechanism of the radio spikes implies the necessity of many e-folding amplifications of the radiation as it propagates through the source region of a given spike. It is apparent that the spike spectrum has a peak where the corresponding spatial growth rate $\varkappa_{\sigma}$ has a maximum as a function of frequency \citep{Fl_2004_AL}. Since for the required strong amplification, only a small vicinity of this maximum plays a role, the true frequency dependence of the growth rate can be reliably approximated by a parabolic one \citep{Fl_2004}
\begin{equation}\label{kappa_parab}
  \varkappa_{\sigma} \simeq \varkappa_{\sigma}^m (1-\alpha
  (f-f_0)^2),
\end{equation}
where $\varkappa_{\sigma}^m$ and $\alpha$ are constant parameters of this approximation. The intensity of the amplified waves is proportional to $\exp(\varkappa_{\sigma}L)$, where $L$ is the spike source size along the line of sight; thus, the parabolic dependence of the growth rate translates to the Gaussian dependence of the wave energy density. The ECM mechanism, assumed here, is direct amplification of the escaping electromagnetic waves, which immediately leads to a Gaussian spectral shape of elementary spike emission.

\subsection{Algorithm description}

\begin{figure*}
\epsscale{0.5}\plotone{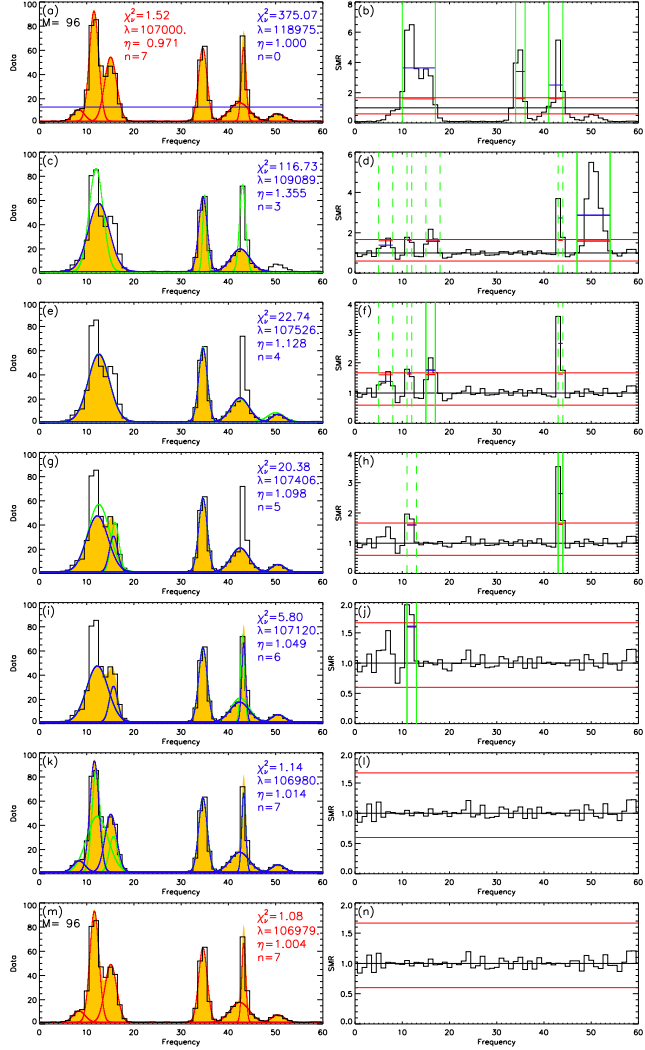}
\caption{\label{fitsteps} Fitting algorithm steps in the  case of a synthetic Gaussian superposition data (panel a--red curves) contaminated by $M=96$ Gamma noise. The final solution, and its ratio to the true data are shown in panels $(m)$ and $(n)$, respectively. At each intermediate step, the fitted Gaussian peaks (blue lines), their corresponding envelope (yellow filling), and the new added component (green lines) are indicated on the left-hand panels. The right-hand panels display the SMR estimator (black lines), the range of the compact regions above unity (vertical green dashed lines), the observed SMR means (horizontal blue segments), their Pearson Type I upper limits (red horizontal segments), and the selected least-probable region where a model adjustment will be attempted in the next step (vertical green solid lines). For reference, the horizontal red lines indicate the Pearson Type I approximation limited range corresponding to an $n=1; M=96$ SMR region. The inset text on the left-hand panels give goodness-of-fit $\chi_\nu^2$, the negative log--likelihood $\lambda$, the averaged SMR, $\eta$, and the current number of spikes $n$ for the evolving fitting solution.}
\end{figure*}

Figure \ref{fitsteps} offers a step-by-step illustration of the algorithm for the case of a synthetic spectrum that contains a superposition of 7 peaks with different amplitudes, bandwidths, and degrees of overlap. A noise level corresponding to $M=96$ has been applied to match the typical accumulation length of the FST instrument \citep{FST} and $N=61$ frequency channels have been chosen to provide an appropriate frequency scale visualizing the spectral details.

\subsubsection{Initial guess}

The first step of the algorithm performs an initial guess of a minimal number of peaks present in the spectrum. The algorithm starts with a simple flat spectrum indicated by the blue horizontal line in Figure \ref{fitsteps}a, which results in the parameters listed in blue on the right side of the plot. Based on this initial model function, the local SMR deviations are computed and displayed in Figure \ref{fitsteps}b. At this stage, the SMR sequence is divided into adjacent compact regions deviating below or above unity, and their mean SMR  deviations $\eta_b$, or $\eta_a$, are computed. The probabilities of observing these averaged SMR deviations are computed according to Eqn. \ref{betaprob} and all SMR regions above the adopted probability threshold are flagged.  An initial guess of Gaussian parameters are computed based on the number, location, width, and average amplitude in the region. For the particular example used for illustration, Figure \ref{fitsteps}b indicates three such regions delimited by solid green vertical lines. Their averaged SMR values are shown as blue horizontal segments, while the red solid segments indicate the maximum range of the deviations allowed by the Pearson Type I PDF approximations in each case. The two red horizontal lines spanning the entire frequency range indicate the standard $0.13499\%$ PFA thresholds corresponding to an $n=1$ region. The set of these three Gaussian peaks shown in Figure \ref{fitsteps}c in solid green along with the flat background are then used as the initial guess of the true spectral shape.

\subsubsection{Step-by-step $\chi^2$ minimization and model adjustment}

The next stage of the fitting algorithm involves a standard $\chi^{2(II)}$ minimization that, in our example, results in the set of three Gaussian peaks shown in Figure \ref{fitsteps}c as blue solid lines. In this and all subsequent left-hand panels, the $\chi_\nu^2$, $\lambda$, and $\eta$ parameters from the fit in that step are indicated in blue. Note that for overlapping peaks the $\chi^{2(II)}$ minimization finds a solution close to the low-amplitude spectral components. This built-in bias of the $\chi^{2(II)}$ minimization, which results in the recomputed SMR deviations shown in Figure \ref{fitsteps}d, offers greater sensitivity for finding additional components as discussed earlier.  Now five regions (vertical green lines) are identified, where new spectral components should be added to the model function.

At this stage of the algorithm, the flagged SMR regions are sorted in order of their chance probability, equation~(\ref{absolute}), and the most improbable one, marked by solid vertical lines in our example, is chosen as the location at which an additional spectral peak is added to the model. Panels $e$-$l$ display a sequence of identical steps that are repeated up to the point that, as seen in panel $l$, no compact SMR region deviates above the upper limit of the Pearson Type I PDF, and no individual SMR deviation, whether isolated or part of a compact region, crosses the $n=1$ standard probability threshold.

\subsubsection{Maximum likelihood correction of the $\chi^2$ minimization}

The last stage of the fitting algorithm consists of using the final solution of the $\chi^{2(II)}$ minimization as a starting point for finding the absolute minimum of the negative log-likelihood function $\lambda$, which, as shown in Figure~\ref{twospikes} (see also Figure~\ref{one}), should be situated in the neighborhood of the $\chi^{2(II)}$ solution. The spectral shape solution provided by the $\lambda$ minimization is displayed in Figure~\ref{fitsteps}m and its corresponding SMR deviations are displayed in Figure \ref{fitsteps}n. One should note that the final $\lambda$ minimization results not only, as expected, in a smaller value of $\lambda$, but also in a smaller value of the goodness-of-fit parameter $\chi_\nu^2$ (which corresponds to the $\chi^{2(I)}$ measure).

Although, in the case of the particular example illustrated in Figure \ref{fitsteps}, this final $\lambda$  minimization stage of the algorithm results only in a minimal change in the final $\chi^{2(II)}$ solution, this correction has the merit of having a proper goodness-of-fit estimate and also being fully consistent with a maximum likelihood criterion.

\subsubsection{Avoiding excessive augmentation of the fitting model}

Each spectral model component has its own amplitude-dependent level of statistical fluctuations; thus, the successful fitting algorithm must fit both the true signal (deterministically) and the associated fluctuations (statistically). Adding spectral components to an already well-determined model will reduce the fluctuations below the expected level, which offers an objective means for rejecting such an excessive model. To avoid adding an excessive spectral component, our fitting algorithm employs a gradual augmentation of the model function, which except for the starting point, adds only one spectral component at each step up to the point where no SMR region exceeds the preset PFA thresholds. Therefore, by design, the algorithm is not expected to continue adding spectral components above the minimally needed number.

However, the $\chi^{2(II)}$ minimization might not find a valid solution for a given set of initial guess parameters, or might settle on a local rather than a global minimum. To avoid this, we implemented a mechanism that checks the validity of solution and, in case of failure, allows an early termination of the process before all SMR residuals are brought within the desired limits. For this purpose, after each new minimization, the goodness-of-fit $\chi_\nu^2$ is computed and the new solution is rejected if its goodness of fit is larger than the one of the previous solution. If this happens, while more than one SMR region had been flagged at the previous step, the next in line is considered, and a new $\chi^{2(II)}$ minimization is attempted based on the new set of initial parameters until the $\chi_\nu^2$ is successfully decreased or all the available flagged regions have been tried. If the attempt to add a new spectral component fails for all SMR suggested locations, the $\chi^{2(II)}$ minimization stage is abnormally exited, and the partial solution may be either rejected or flagged as possibly unreliable.

\subsection{Influence of spectral resolution on algorithm performance}

\begin{figure*}
\epsscale{0.85}\plotone{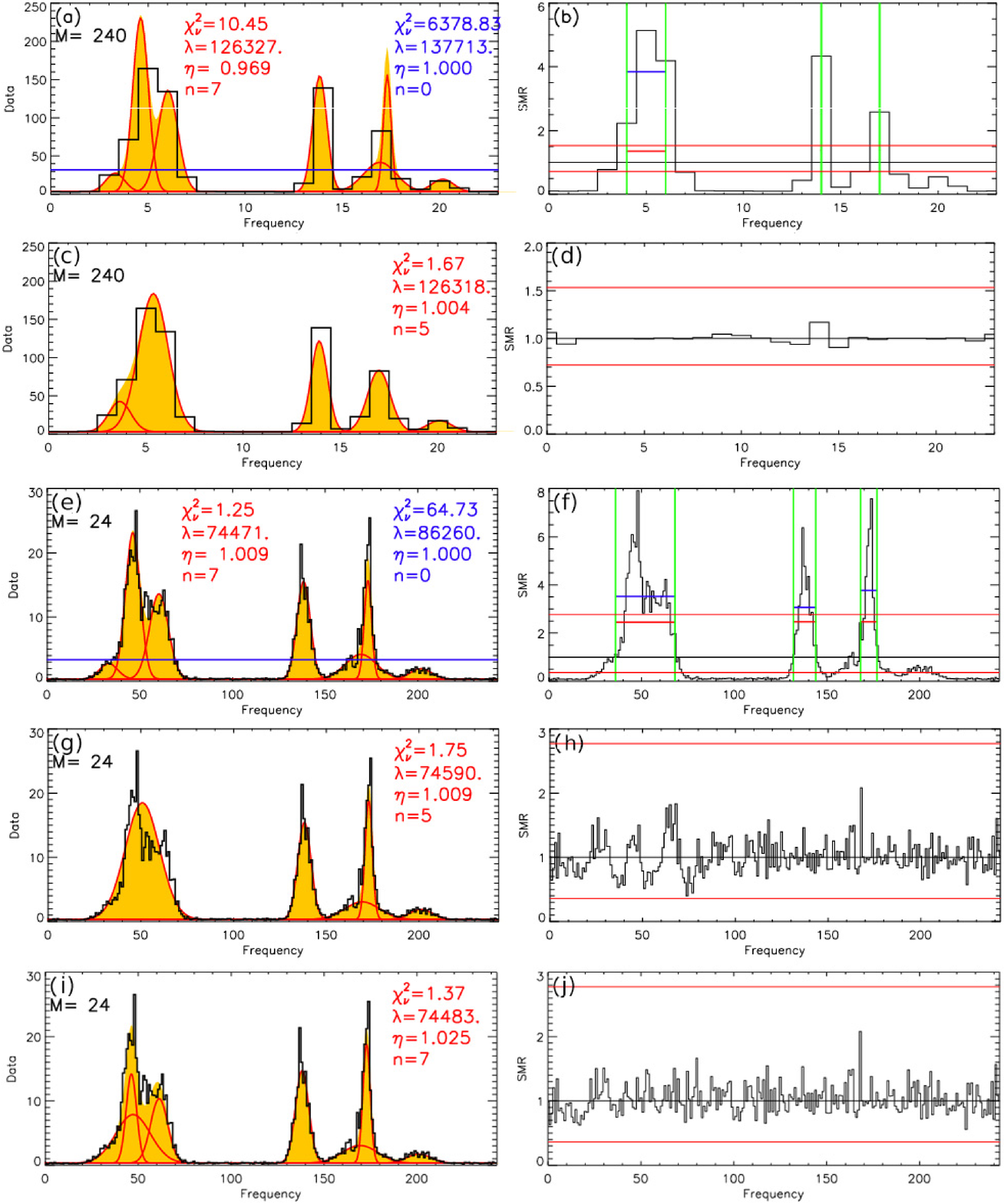}
\caption{\label{tf_res} (a-d) First and last steps of the fitting algorithm for the same synthetic Gaussian superposition as in Figure~\ref{fitsteps}, as would be seen with a larger number of accumulations ($M=240$), which is achieved at the cost of decreasing the frequency resolution. The noise variance is reduced by a factor of $2.5$, but the lower frequency resolution does not fully resolve all of the spectral features in the model. (e-h) First and last steps of the fitting algorithm for the same synthetic Gaussian superposition, as would be seen with higher frequency resolution and a smaller number of accumulations ($M=24$), which increases the noise variance by a factor of $4$. (i-j) The last step of the fitting algorithm corresponding to (g-h), but for a tiny increase in PFA threshold of $10^{-12}\%$, which now succeeds in separating the spike cluster near frequency bin 50 into its three components.}
\end{figure*}

As noted earlier, the FST instrument directly records time-domain data for 100~$\mu$s ($10^5$ samples) at a time.  These can be split into $M$ accumulations of $N$-channel spectra, with the constraint $2MN = 10^5$, allowing us to trade spectral resolution, 500/$N$ MHz, for a reduced level of statistical fluctuations, which depend on $M$. This motivates us to consider the effect of this trade-off, using simulated data, on the algorithm's ability to recover the true model.
Figure \ref{tf_res} illustrates the algorithm's performance in the case of a synthetic spectrum composed of the same underlying components as in Figure \ref{fitsteps}, but sampled with lower and higher spectral resolutions. To mimic the design of an FFT-based spectrometer like FST, which digitizes the time domain signal at a fixed sampling rate, the changes in spectral resolution are accompanied by inversely proportional changes in the accumulation lengths $M$.

The spectrum shown in Figure \ref{tf_res}a,c, although less affected by noise fluctuations due to its increased accumulation length, $M=240$, is negatively impacted by the proportionately lower spectral resolution. It results in under-sampling the spectral shape, as quantified by the large $\chi_\nu^2=10.45$. In these conditions, the algorithm, while succeeding in reducing the SMR deviations, fails to properly resolve the overlapping spectral peaks, and underestimates the amplitudes of the isolated ones, even though the final $\chi_\nu^2=1.45$ suggests a rather successful fit.
On the other hand, the higher frequency resolution of the spectrum shown in Figure~\ref{tf_res}e,g, which should facilitate the spectral separation of the overlapping peaks, is counterbalanced by the shorter accumulation length $M=24$, which results in a higher level of statistical fluctuations. This level of fluctuations prevents the algorithm from resolving three closely overlapping peaks. However, the local systematic SMR deviations revealed by Figure \ref{tf_res}h, suggests the need for further model adjustment. To programmatically achieve this, the detection threshold would have to be decreased at the cost of a higher false-alarm rate. Figure \ref{tf_res}i presents the last stage of the fitting process of the same spectrum as in Figure \ref{tf_res}g, but with the choice of a tiny increase in the PFA threshold of $10^{-12}\%$. The algorithm now succeeds in lowering the local SMR deviations and identifying the correct number of spectral peaks.

\section{Analysis of the 06 December 2006 19:41:00 UT data recorded by FST}
\label{data}

We now apply the fitting algorithm to a 60-s data segment recorded by the FST instrument \citep{FST} during the record-breaking solar flare of 06 December 2006 \citep{Cerruti}. During this exceptional solar flare, which lasted more than an hour, the radio data recorded by the FST instrument in the $1.0-1.5$GHz frequency range revealed a rich variety of spectral fine structures such as fiber, zebra, and spike bursts. The data that we analyze occurred at the decay phase of a much longer period of spike emission that was mainly responsible for the highest radio flux ($>10^6$~sfu) ever recorded during a solar radio burst.  The spikes became less numerous during this late decay phase, but nevertheless occurred in such high numbers that many overlapping spikes are seen, which the algorithm described above must handle.

\begin{figure*}
\epsscale{1}\plotone{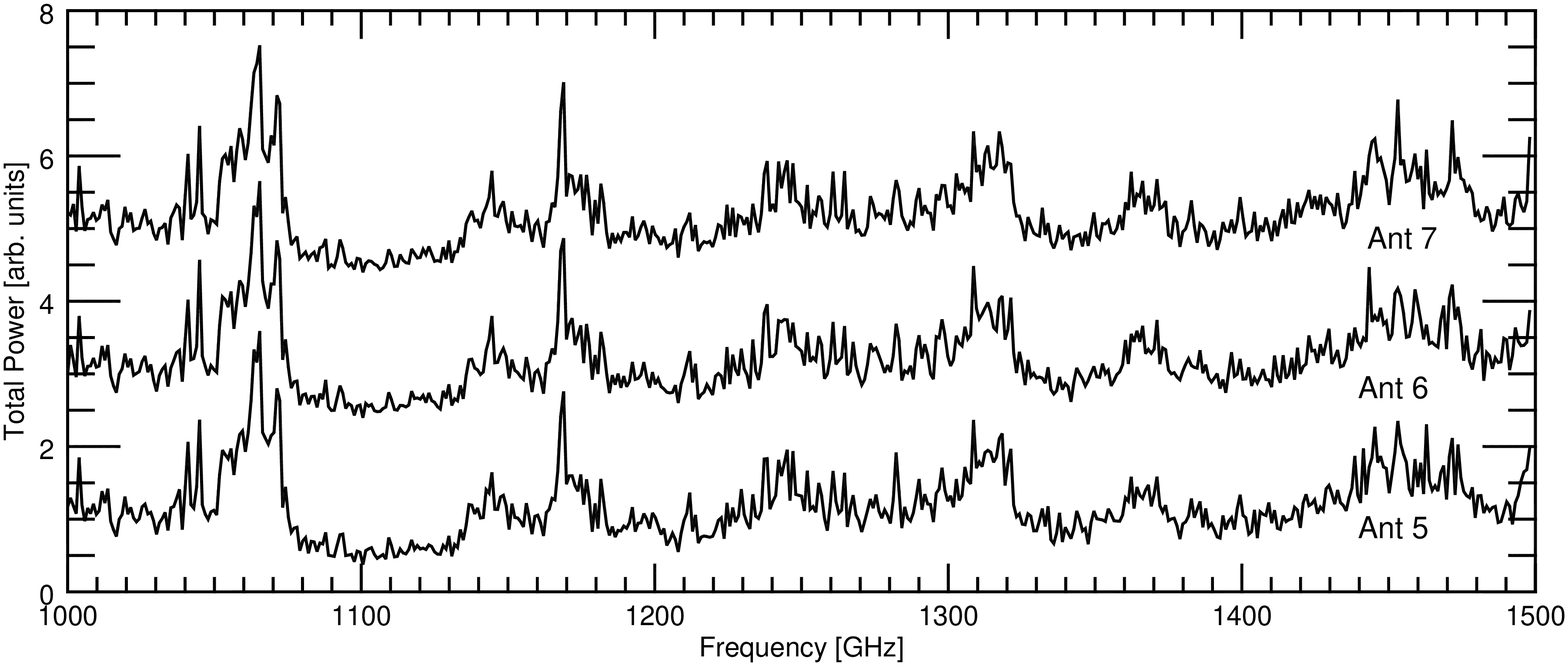}
\caption{\label{TP_compare} Comparison of total power records for each of the three antennas (5, 6 and 7) for a single time as a function of frequency, after normalizing to the background continuum. The signal for antenna 6 is shifted by 2, and the signal for antenna 7 is shifted by 4 in amplitude for clarity.  The signals are essentially identical down to even minor fluctuations, especially at the lower end of the frequency range where the continuum is stronger.}
\end{figure*}

FST observes the Sun interferometrically with three antennas.  Although both total power (signals from each antenna separately) and correlated data (cross-correlations of amplitude and phase between each pair of antennas) are available, the FFT-based statistics we describe in this paper apply only to total power. As shown in Figure~\ref{TP_compare}, the total power records are essentially identical, which confirms that the fluctuations are dominated by the statistics of the solar signal and not by uncorrelated thermal fluctuations in each antenna.  Incidentally, it also places a constraint on the anisotropy in the spike emission on the scale of the spatial separation between the antennas (baselines of 136, 244, and 280 meters): there is clear one-to-one correspondence between spikes recorded by the three different antennas in contrast to what has been reported by \cite{Dabrowski_etal_2011} for two different instruments located in Toru{\'n} and Ond{\v r}ejov and so separated by 450~km.

\begin{figure*}
\epsscale{1}\plotone{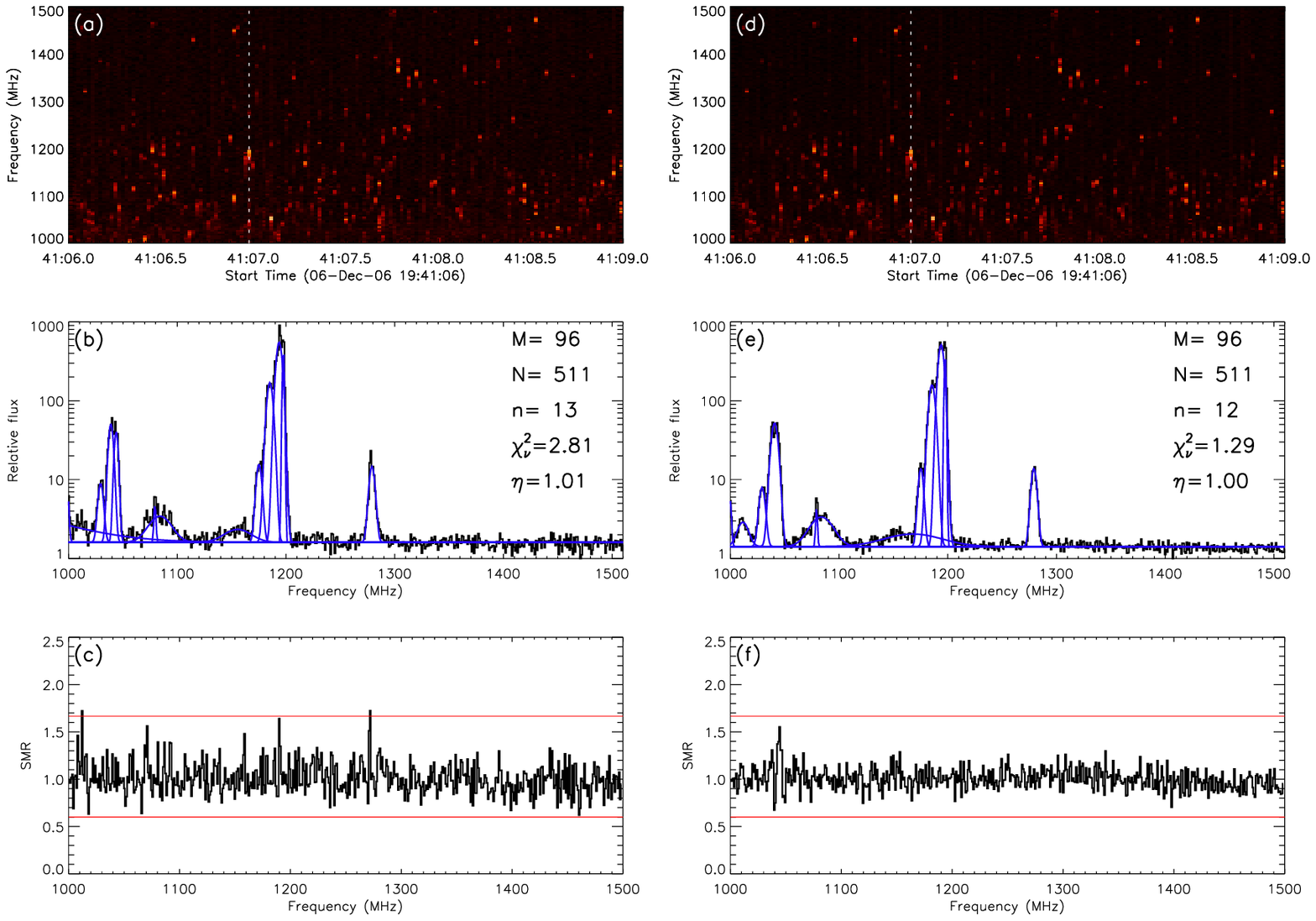}
\caption{\label{fst} a) Dynamic spectrum segment of the 06 December 2006 solar radio burst data recorded by FST in the $1-1.5$GHz frequency range with $N=511$ frequency channels ($\sim$1MHz resolution) and an $M=96$ accumulation length. The abundant spike emission displaying different degrees of overlapping is evident. The dotted vertical line indicates a selected time frame used for illustration. b) Data and estimated spectral components for the selected time frame. The estimated number of spikes, $n=13$, the goodness of fit $\chi_\nu^2=2.81$, and the averaged SMR deviations $\eta=1.01$ are indicated in the figure inset. c) Local SMR deviations corresponding to the solution shown in panel (b). All but two SMR deviations lie within the maximum allowed range ($0\%$ PFA) of the $n=1$ Pearson Type I PDF. Panels $(d,e,f)$: The same as in panels $(a,b,c)$, but for the synthetic data set built by contaminating the solution estimated from real data with pure $M=96$ statistical noise. }
\end{figure*}

Figure \ref{fst}a displays a 3-second detail of the $M=96$ dynamic spectrum showing the spike emission. The vertical dotted line indicates the time of a representative fit solution shown in Figure~\ref{fst}b. The $\chi_\nu^2$ goodness-of-fit parameter, and the average SMR deviation $\eta$ are printed in the plot. Figure \ref{fst}c displays the local SMR deviations. All but two local SMR deviations lie within the $0\%$  PFA thresholds shown by the horizontal lines, but the algorithm did not add spikes in these locations because doing so did not improve $\chi_\nu^2$.

{We can check to what extent the model derived from the data is consistent with our expectation that the statistical noise is dependent solely on the amplitude of the signal.  To do this, we generate a synthetic dynamic} spectrum by taking the superposition of Gaussians {found for} each spectrum of real data and adding numerically generated statistical noise calculated for $M=96$. The resulting {synthetic} spectrum for the 3-s period is shown in Figure~\ref{fst}d, {which cannot be distinguished visually from Figure~\ref{fst}a. We then run the algorithm on these synthetic data and examine the fit results and the residuals.  For example,} the fitting solution and corresponding SMR deviations for the same time as in Figure~\ref{fst}b,c are shown in Figure~\ref{fst}e,f, respectively. Although qualitatively similar to the real-data solution, the synthetic-data solution in Figure~\ref{fst}e has only $n=12$ spikes and a better $\chi_\nu^2=1.29$. The SMR residuals in Figure~\ref{fst}c appear to be about twice those of Figure~\ref{fst}f. This quantitative mismatch between the data-based and simulation-based solution indicates that not all SMR residuals in Figure~\ref{fst}c are due to $M=96$ noise.

\begin{figure*}
\epsscale{1}\plotone{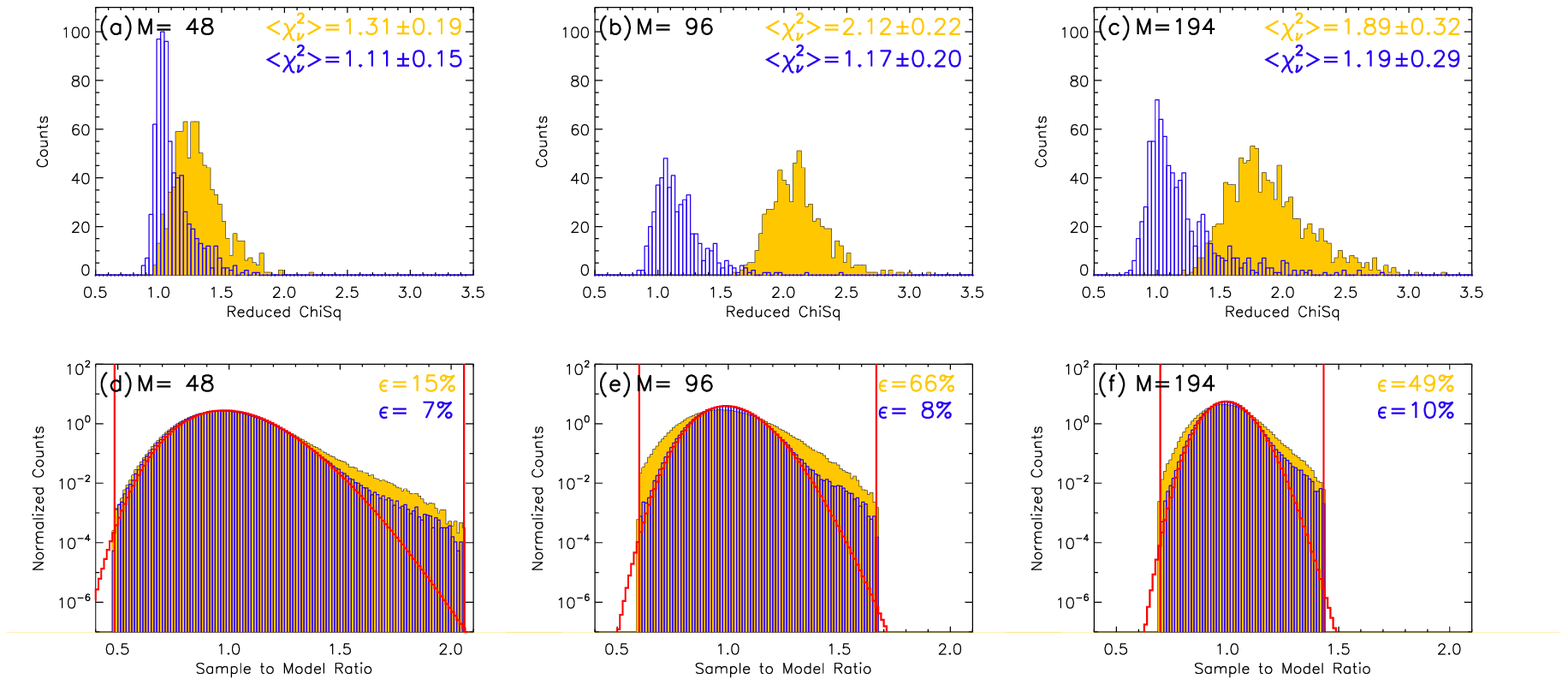}
\caption{\label{gof} Goodness-of-fit analysis for the 150 spectra in Figure~\ref{fst}a,d, obtained with three different accumulation lengths, $M=48, 96$ and 194. Upper panels: $\chi_\nu^2$ distributions for the original data (solid yellow) and for synthetic data (blue lines). The means and standard deviations of the $\chi_\nu^2$ parameters are indicated on each plot. Lower panels: Distributions of the SMR deviations for the original (solid yellow) and synthetic (blue lines) data. The distributions theoretically expected according to equation~(\ref{normpdf}) are plotted in each panel (red lines) and the percentage $\epsilon$ of points falling outside the theoretical distribution is shown on each plot. Only {those time frames for which the algorithm has found solutions} with all SMR deviations ranging within the expected theoretical limits, i.e. the $0\%$ PFA thresholds indicated by vertical red lines, were selected for this analysis.}
\end{figure*}

This is further confirmed by Figure~\ref{gof}, which presents the distribution of $\chi_\nu^2$ for the 150 times in Figure~\ref{fst}a,d and the distributions of the SMR deviations corresponding to each spectral point of these times. The analysis is performed for three different accumulation lengths, $M=48$ (left), $M=96$ (middle), and $M=194$ (right). Here we include only those times for which the fitting algorithm succeeded to find a solution where all SMR deviations are within their allowed range ($0\%$ PFA).  Thus, for example, the fit in Figure~\ref{fst}b is rejected.

{The $\chi_\nu^2$ distributions for the synthetic spectra in Figure~\ref{gof}a-c cluster around unity, while those based on the original data are higher.  This suggests that there is another source of fluctuations than those based on accumulation number $M$, which becomes masked only when $M$ is small enough that $M$-based fluctuations dominate. These fluctuations could be due to unresolved spikes.
The distributions of the individual SMR deviations shown in the bottom row of Figure \ref{gof} complement the above conclusion by revealing that the SMR distributions of both real (solid yellow) and synthetic (blue) data fits are not only closest to the theoretical distribution (red curve) in the $M=48$, but also consistent to each other, i.e. they have similar excess values $\epsilon=15\%$ and $\epsilon=7\%$, respectively, relative to the expected theoretical SMR distribution given by equation (\ref{normpdf}).}

{
For the remainder of this section we continue to compare the results derived from actual data in yellow with those derived from model-based synthetic spectra in blue, as a means of highlighting potential limitations.}

\subsection{Statistical distributions of spike parameters}

\begin{figure*}
\epsscale{1}\plotone{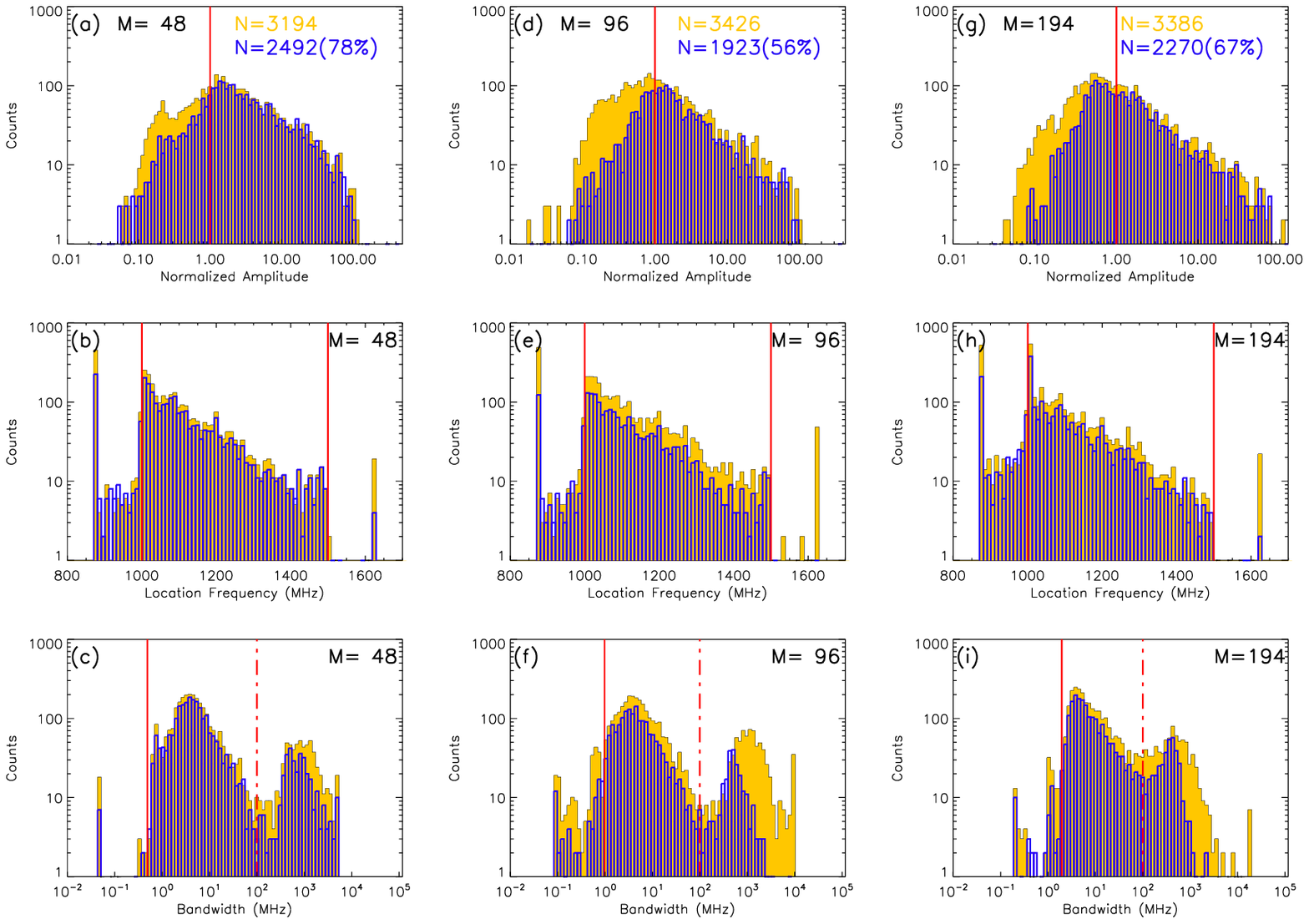}
\caption{\label{real2sim} {Upper row: Normalized amplitude distributions. Middle row: Central frequency distributions. Bottom row: Bandwidth distributions. The first, second, and third row display the distributions corresponding to $M=46$, $M=96$ and $M=194$, respectively. In all panels, the solid yellow histograms correspond to the spikes extracted from real data, while the blue histograms correspond to the spikes extracted from synthetic data generated based on the previously extracted spikes. For reference, we indicate in the upper row plots the normalized unity amplitude by vertical red solid lines. In the middle row plots, the pairs of vertical red solid lines mark the physical boundaries of the observed $1-1.5$ GHz frequency range. In the bottom row plots, the vertical red solid lines indicate the spectral resolutions corresponding to each accumulation length, while the dashed red vertical lines mark the 100 MHz boundary empirically found to separate two apparently distinct components of the bandwidth distributions.} }
\end{figure*}

Figure \ref{real2sim} compares the data-derived distributions (solid yellow histograms) of spike normalized amplitude (upper row), location frequency (middle row), and bandwidth (bottom row), with the distributions derived from the corresponding synthetic spectra (overlaid blue histograms).

The amplitude normalization factor has been chosen based on the average FST background in absence of any spike emission. For reference, the unit normalized amplitude, which therefore correspond to the average noise level, is indicated on each of the upper panel plots by vertical red lines. The number of spikes extracted from real data, i.e. $N=3194(M=48)$, $N=3426(M=96)$, and $N=3386 (M=194)$, as well as the number and relative percentage of the extracted spikes from the synthetic spectra, i.e. $78\%(M=48)$, $56\%(M=96)$, and $67\%(M=194)$, are shown in the plots. The amplitude distributions display high-end power-laws, as well as a low-end rollover clearly associated with the background noise level. While the power-law slopes and intercepts of the real- and synthetic-data amplitude distributions are consistent in each case, it is evident that most of the missing spikes from the synthetic-data distributions come from below the noise level. We conclude that the distributions above unit normalized amplitude are reliable, and further, as we concluded from Figure~\ref{gof}, the $M=48$ spectra provide the best combination of statistical noise and frequency resolution. The spike extraction algorithm performance can be quantified by an overall $78\%$ real-spike validation rate, as well as by only a minor difference between the lower ends of the real and synthetic data distributions.

The center-frequency distributions of the spikes, shown in the middle row of Figure~\ref{real2sim}, have a remarkably pure exponential distribution in the available spectral domain, 1---1.5~GHz. The shape of the distribution, however, suggests that there are spikes at $f>1.5$~GHz and many more at $f<1$~GHz. We cannot know the overall frequency distribution including these unmeasured bands.

The bandwidth distributions shown in bottom row of Figure~\ref{real2sim} show a bi-modal distribution, but the broader-band population centered around 1000~MHz exceeds the 500~MHz bandwidth of the observed spectrum. These large bandwidth structures are added by the fitting algorithm to compensate for smooth variations of the background, and are not actual spike emission. The lower-bandwidth distribution, from a few to about 100~MHz (vertical red dotted line), show a clear power-law pattern with a roll-over below a few MHz.  This appears to be a real attribute of the spike bandwidth distribution, since it is clearly located above the corresponding spectral resolution, i.e. 0.5, 1, and 2~MHz, indicated in each case by red vertical lines in Figure~\ref{real2sim}c, f and i, respectively.

\begin{figure*}
\epsscale{1}\plotone{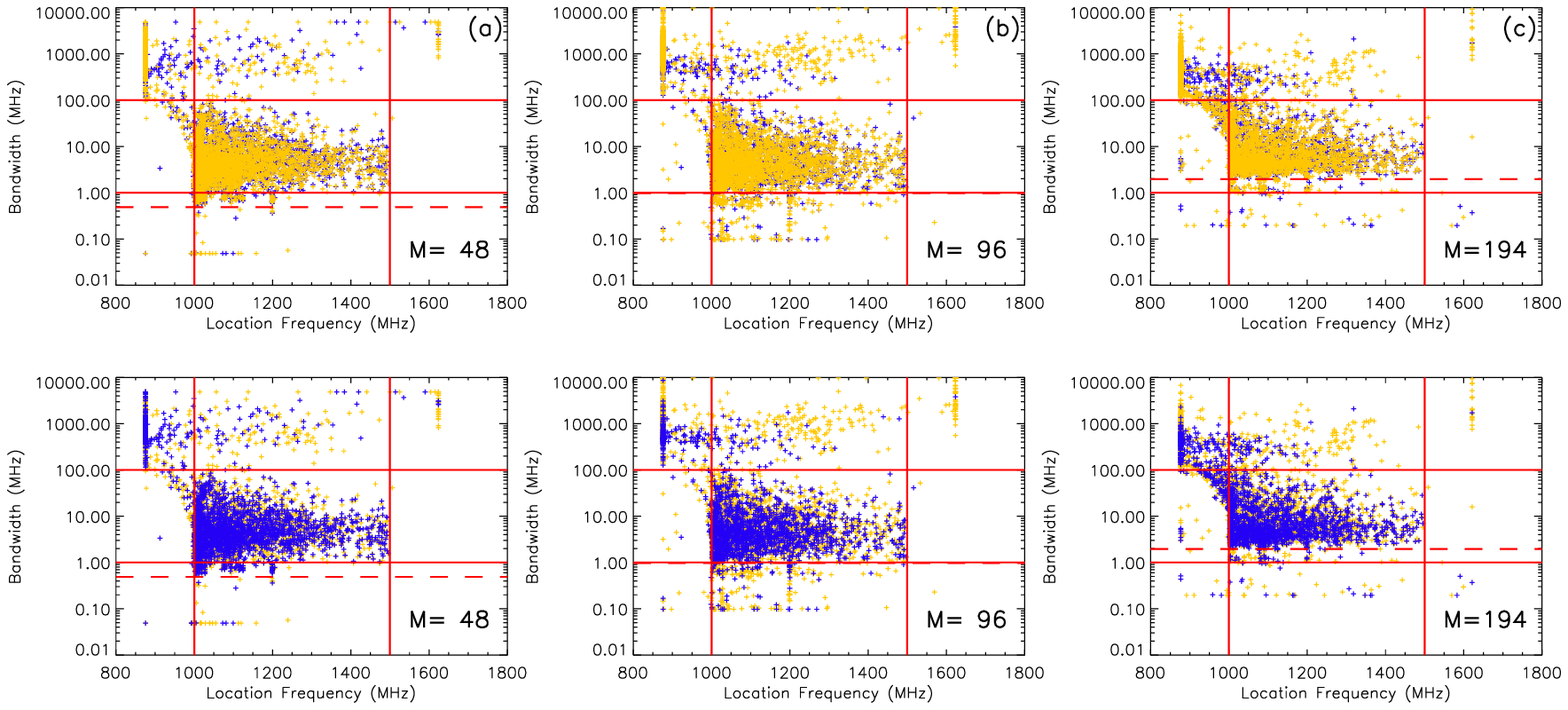}
\caption{\label{freq2bw} Two dimensional distributions of spike locations versus spike bandwidth. The upper and bottom panels alternate the order in which spikes extracted from the original (yellow) and synthetic (blue) data are plotted on top of each other. {The pair of solid red vertical lines represent the $1-1.5~GHz$ frequency range of the instrument, the pair of solid red horizontal lines represent a fixed $1-100~MHz$ bandwidth range, and the dotted red horizontal line indicates the frequency resolution corresponding to the accumulation length $M$. }}
\end{figure*}

In order to further distinguish solar spikes from non-solar artifacts, we present in Figure~\ref{freq2bw}, for each of the three accumulation lengths, the two-dimensional distributions of spike location versus spike bandwidth. For clarity we plot in the top row the real-data spike distribution (yellow) on top of the synthetic-data distribution (blue), while in the bottom we invert the plotting order. The pair of solid red vertical lines represent the $1-1.5~GHz$ frequency range of the instrument, the pair of solid red horizontal lines represent a fixed $1-100~MHz$ bandwidth range, and the dotted red horizontal line indicates the frequency resolution corresponding to the accumulation length $M$. There is a clear separation of fitted spikes into two families, one within the physically justifiable confines of the inner red box, and another well outside the region of interest. Therefore, we conclude that the parameter-space boundaries drawn by the red solid lines, along with a minimum normalized-amplitude limit of unity, can be used as filtering criteria to reliably separate the true solar spikes from non-solar artifacts.

\begin{figure*}
\epsscale{1}\plotone{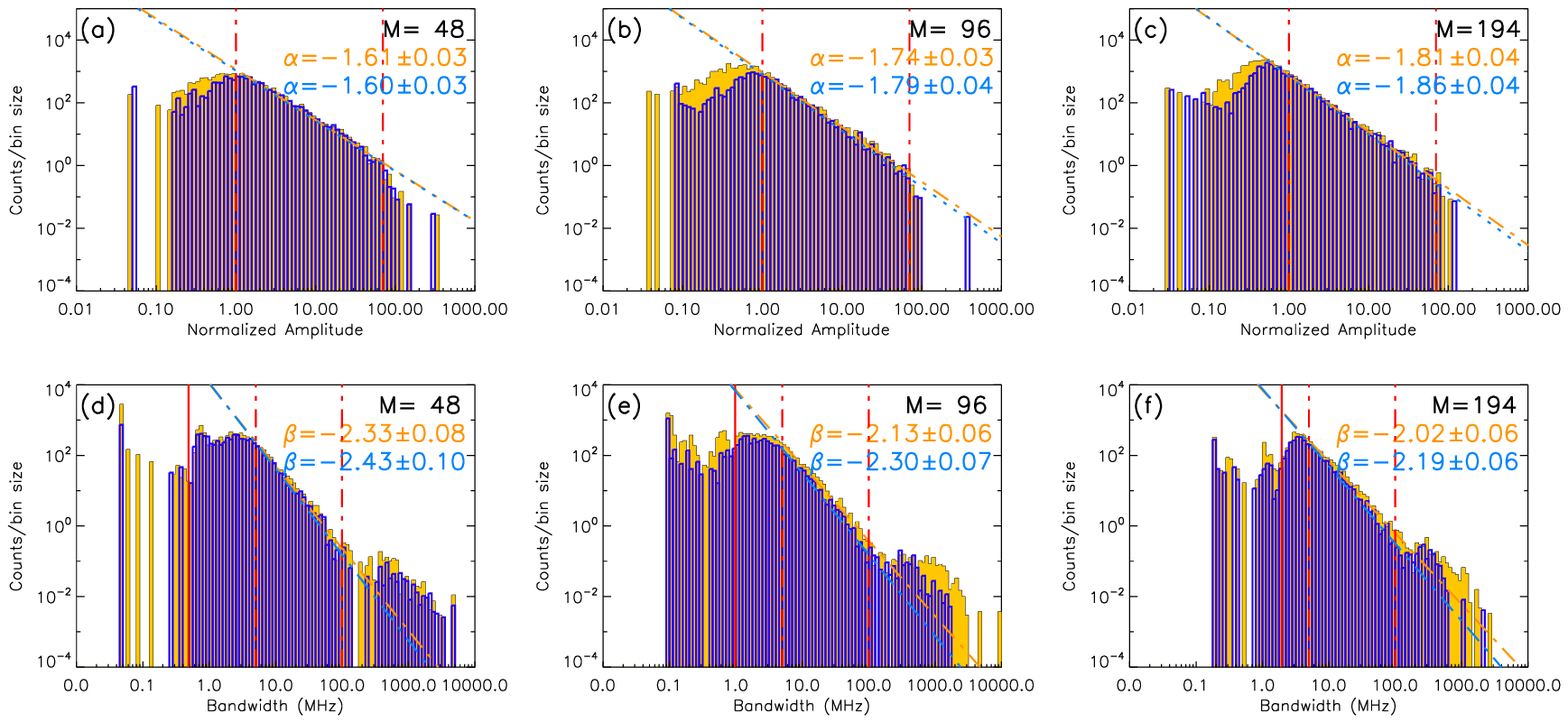}
\caption{\label{ampbw} Power-law fitting of amplitude and bandwidth distributions derived from real (yellow) and synthetic (blue) data. {The amplitude distributions displayed on the upper row were filtered based on the bandwidth and location frequency boundaries shown on Figure \ref{freq2bw}, while the bandwidth distributions shown on the bottom row were filtered based on the location frequency criterion, as well as by the minimum unit relative amplitude criterion.} The fitting ranges, {which complete the 3--criterion filtering process} are indicated by vertical red dash-dotted lines and the power-law indexes and their associated fit uncertainties of real and synthetic distributions are indicated in corresponding colors on each plot. {On the bottom row plots, the vertical red solid lines indicate the spectral resolution corresponding to each accumulation length $M$.}}

\end{figure*}

Consequently, Figure \ref{ampbw} presents the results of fitting the power-law indexes of the filtered amplitude and bandwidth distributions. {However, for illustration purposes, the distributions shown in this figure were only partially filtered as follows: to build the density distributions shown on the upper row, only the bandwidth and location frequency filters were applied, while the bandwidth distributions shown on the bottom row were obtained from data filtered only by the location frequency and amplitude criteria.} In contrast to Figure \ref{real2sim}, the distributions shown in Figure \ref{ampbw} are the density distributions, i.e. the counts in each logarithmic bin have been divided by the corresponding variable bin width. On each plot, the amplitude or bandwidth ranges used to perform the linear fits{, which complete in each case the full filtering of data,} are indicated by vertical red dotted lines. The resulting power-law indexes, and their fit uncertainties, are indicated in corresponding colors for the real (yellow) and synthetic (blue) distributions. {On the bottom row plots, the vertical red solid lines indicate the spectral resolution corresponding to each accumulation length $M$.}

The results displayed in Figure \ref{ampbw} confirm quantitatively that the best agreement between the slopes of the real and synthetic distributions is reached in the case $M=48$, the most favorable tradeoff between spectral resolution and the expected level of statistical noise fluctuations for the solar spikes observed in this event.

\begin{figure*}
\epsscale{1}\plotone{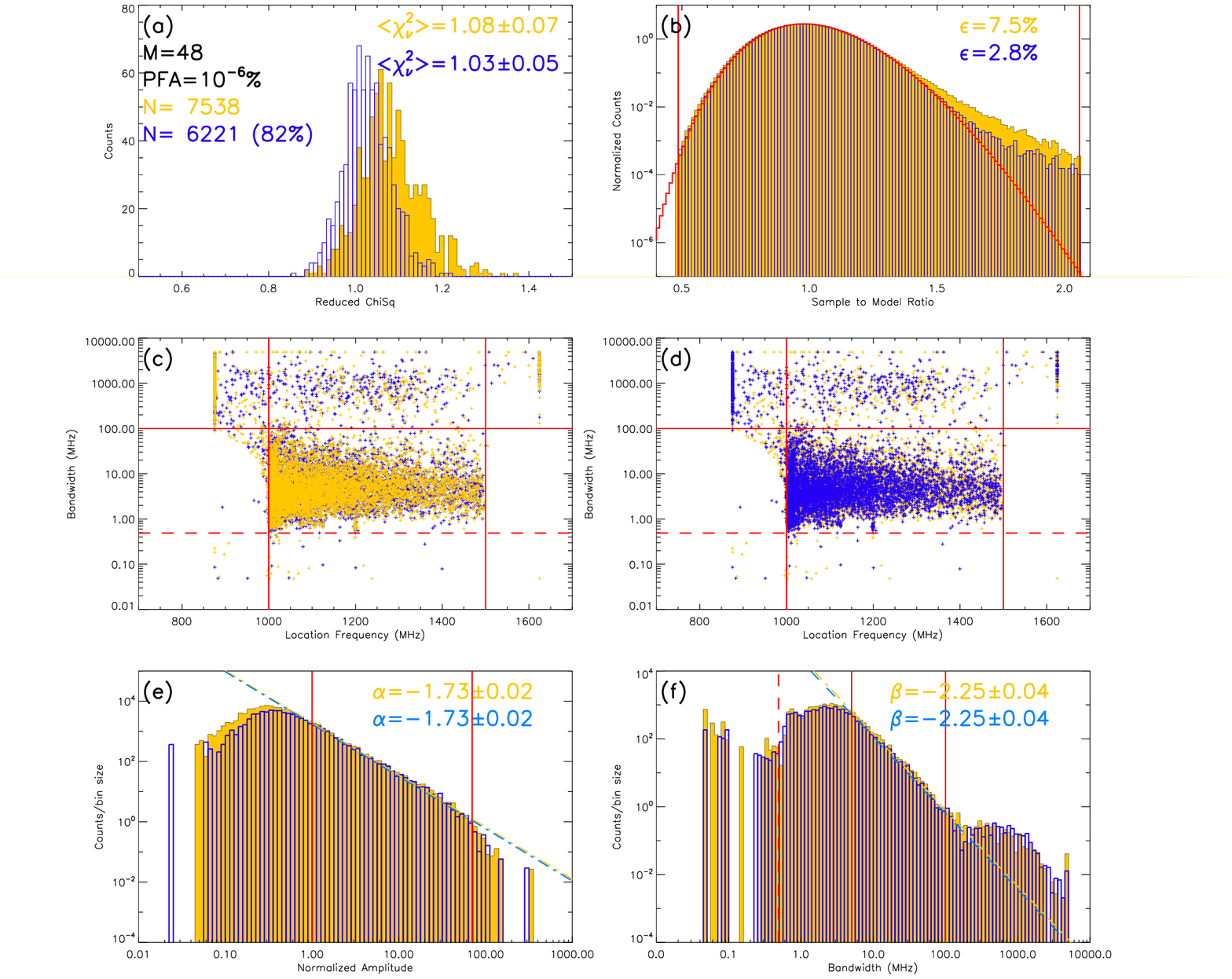}
\caption{\label{selected_case} {Characteristics of the extracted spikes from the $M=48$ spectrum with $p=10^{-6}\%$ PFA as obtained from the original (yellow) and corresponding synthetic (blue) data. Upper row: Goodness of fit analysis presented in panels (a) and (b) in the same format as in Figure  \ref{gof}. Middle row: Two-dimensional distributions of spike locations versus spike bandwidth, as in Figure \ref{freq2bw}. Panels (c) and (d) present the same information with alternate order in which spikes extracted from the original (yellow) and synthetic (blue) data are plotted on top of each other. Bottom row: The amplitude (panel e) and bandwidth (panel f) distributions, and their corresponding power-law fits, as in Figure \ref{ampbw}.}}
\end{figure*}

As one final refinement, we noted earlier (Fig.~\ref{tf_res}) that the finding of spikes might be improved by lowering the spike detection thresholds from the $PFA=0\%$ level to a less conservative, but still close to zero, $PFA=10^{-6}\%$. We therefore repeat the fitting based on this detection threshold for $M=48$ and summarize the results in Figure~\ref{selected_case}. In Figure~\ref{selected_case}a we compare the goodness of fit estimators of the original-data fits, $\langle\chi_\nu^2\rangle=1.08\pm0.07$, with those from fitting the corresponding synthesized spectra, $\langle\chi_\nu^2\rangle=1.03\pm0.05$. Changing the PFA level to a small, but a nonzero value resulted in more than doubling the number of spikes extracted from data, i.e. $N=7538$ vs $N=3194$, while the percentage of self-consistently validated spikes increased slightly to $82\%$ from {the} $79\%$ {figure} indicated in Figure\ref{real2sim}a. We interpret this {result} as an indication that using a slightly less conservative PFA threshold allows extraction of significantly more spikes, and consequently a better spike model, {without a significant increase} in misidentification of statistical noise fluctuations as true solar spikes. {To quantitatively support this assertion, we calculate that, when directly applied to the number of extracted spikes, $N=7538$, the particular value we chose for the probability for a given detected peak to be created by a statistical fluctuation, $PFA=10^{-6}\%$, indicates that the probability to have any one misidentified spike from the detected $7538$ peaks is $7.538\times 10^{-3}$; hence, from a statistical perspective, less than one additionally detected spike could have resulted from a false alarm flagging of the spectral model. We thus conclude that, for practical purposes, the criterion of keeping the absolute number of potentially false spikes below 1 may provide an objective means for adoption of a certain non-zero PFA suitable for the amount of data being investigated.}

The distribution of the local SMR deviations shown in Figure~\ref{selected_case}b, when compared with Figure~\ref{gof}d, also indicates an improvement in algorithm's performance, which is quantized by a reduction of the distribution excess parameters from $\epsilon=15\%$ to $\epsilon=7.5\%$ for the real data fits, and from $\epsilon=7\%$ to $\epsilon=2.8\%$ for the synthesized data fits.

Similarly, the two-dimensional distributions of spike bandwidth versus spike location shown in Figures~\ref{selected_case}c,d, confirm that the rectangular parameter space bounded by the limits of the observed frequency range (vertical red solid lines), the spectral resolution (horizontal red dashed line) and the empirical high bandwidth limit of 100~MHz (horizontal red solid line), clearly confines the bulk of the true solar spike population and provides a reliable means for filtering out the non-solar artifacts.

The very good visual agreement of the spike normalized-amplitude and bandwidth distributions obtained from the original data (solid yellow) and synthesized data (blue) in Figure~\ref{selected_case}e,f is quantitatively confirmed by their identical power law slopes, $\alpha=-1.73\pm0.02$ and $\beta=-2.25\pm0.04$. The low-amplitude roll-over in Figure~\ref{selected_case}e is well below the noise (unit normalized amplitude) limit, and hence the shape of the true distribution in this amplitude region is not well recovered, as seen from the mismatch between the yellow and blue distributions. In contrast, the low-bandwidth roll-over (peaking around 3~MHz) in Figure~\ref{selected_case}f is well above the frequency resolution (0.5~MHz) and is undoubtedly a real feature.  The power-law tail of this distribution results in a mean bandwidth (7.5~MHz) that is higher than the peak (most probable) bandwidth.

\section{Discussion}
\label{discussion}

The proposed spike decomposition algorithm allowed us to extract more than 7500 individual spikes from a relatively sparse fragment of the powerful spike cluster observed from the 06 Dec 2006 flare; the obtained number of spikes is sufficient for a detailed statistical analysis of spike properties. The fitted distributions of the spike bandwidth have a (real) primary peak at a few MHz and two secondary peaks that we have concluded are artifacts---one below 1~MHz and the other around 1000~MHz. The first of them originates from very narrowband (unresolved) fluctuations (possibly including interference) misidentified by the algorithm as true spikes while the second is due slight background variations. Filtering these two artifacts results in sharp edges in the distribution that prevent us from straightforwardly finding the moments of the true spike bandwidth distribution.  The reliable portion of the measured bandwidth distribution obeys a power-law with index around $-2.25$, mean of 7.5~MHz, and mode of about 3~MHz.

Similar asymmetric shapes of the bandwidth distribution are typical and have been reported for both decimeter and microwave spikes \citep{Elgaroy_Sveen_1973, Csi_Benz_1993, Messmer_Benz_2000, Rozh_etal_2008, Nita_etal_2008}. \citet{Rozh_etal_2008} employed a theory of the ECM bandwidth formation in a source with random inhomogeneities of the magnetic field developed by \citet{Fl_2004} and demonstrated that the observed distribution was in almost perfect agreement with this theory. The comparison of the moments of the observed and modeled distributions yielded an estimate of the corresponding magnetic irregularities (relatively small-scale turbulence) to be $\left<\delta B^2\right>/B^2 \sim 10^{-7}$. In our case, because of the aforementioned high-end and low-end artifacts, we cannot confidently estimate the higher moments of the bandwidth distribution, which is needed to compute the magnetic irregularity level. We, nevertheless, can tell that a comparably low level of the magnetic irregularities would be sufficient to yield the observed distribution.  Although it may seem surprising that such a small value can in fact be measured through its effect on the spike spectrum shape, one must keep in mind that these small magnetic field variations have to be compared with another very small value---the ECM natural bandwidth, which is typically as small as $\sim0.2\%$ \citep{Fl_2004_AL}.

The central frequency, $f_0$, distribution carries some information of the global parameter range at the source of the spike cluster, presumably, one (or more) coronal loops involved in the flaring process. The available spectral range, 1--1.5~GHz implies the magnetic field range of 357--535~G if the ECM emission is produced at the fundamental of the gyrofrequency, or of 179--268~G in case of the second harmonic. Clearly, the actual range of the magnetic field is much broader because the central frequency distribution continues well outside our spectral 'window'. The ability to produce either fundamental or harmonic ECM emission also constrains the plasma density, because the growth rates at these gyroharmonics have a strong dependence on the `plasma parameter' $Y=\omega_{pe}/\omega_{Be}$ \citep[][and references therein]{Fl_Meln_1998, Stupp_2000}. In particular, fundamental extraordinary-mode ECM emission requires $Y<0.25$ ($n_e\lesssim 8\times10^8$~cm$^{-3}$), fundamental ordinary-mode ECM emission requires $0.25<Y<1$ ($n_e\lesssim 1.2\times10^{10}$~cm$^{-3}$), and second harmonic extraordinary ECM emission requires $1<Y<1.4$ ($n_e\lesssim 6\times10^9$~cm$^{-3}$). Identification of the emission mode (recall the spike emission is 100\% polarized in this event) through the spatially resolved measurements along with other context data can help to identify the emission mode and so improve the use of the spike properties as a probe of source parameters.

The derived distribution of spike amplitude does not display any apparent artifact and so is more conclusive. Indeed, in the meaningful range, roughly between 1 and 100, the distribution is well fitted by a power-law with an index around $-1.6$, while it displays low- and high- amplitude rollovers outside this range. The low-amplitude rollover results from inability of the algorithm to fully recover the peaks having the amplitudes comparable or lower than the mean level of the signal. We emphasize that there must be plenty of the low-amplitude spike for a reason related to the FST recording mode. The FST time resolution is 20~ms; however, the signal accumulation time is only 100~$\mu$s within the 20~ms time interval. We checked that the spike signal level does not change measurably during this 100~$\mu$s time interval implying that the typical spike duration is noticeably longer than 100~$\mu$s. On the other hand, we did not find clear cases when the same spike would be recorded over two consequent 20~ms intervals, implying that the spike duration is shorter than 20~ms; note that according a phenomenologically established regression law, the typical spike duration at 1~GHz is only marginally less than $20$~ms \citep[see Fig. 3 in][]{Rozh_etal_2008}. This means that during one measurement the given spike comes and goes, so the 100~$\mu$s snapshot can correspond to any point of the rise, peak, or decay phase. Even if all the spikes were to have the same peak amplitude, the described measurements would result in a broad distribution of the observed amplitude whose exact shape depends on the exact shape of the spike light curve; for example, for an exponential growth and a similar exponential decay this will be a power-law distribution with index $-1$. The observed index, $-1.6$, deviates from $-1$ and so implies some statistical distribution of spike peak amplitudes. It is, thus, difficult to determine a true amplitude distribution from our data given that we do not know the exact shape of the spike light curve.

In fact, the amplitude (or, alternatively, spike power) distributions have been analyzed for a number of events observed with different instruments \citep{Meszarosova_etal_2000,Rozh_etal_2008, Nita_etal_2008, Benz_etal_2009} because various theories of the spike generation predict distinctly different amlitude/power distributions; specifically, the power-law, log-normal, and exponential distributions have been proposed \citep[e.g.,][]{Benz_etal_2009}. The power-law distribution obtained here appears to be in a remarkable agreement with expectation for an avalanche process in which exponential growth of the unstable waves responsible for the spike generation is terminated at high but random level, \citep[see][and, specifically, their curve b in Figure~4]{Aschwanden_etal_1998Avalanche}.
In our case, the role of this termination of random wave growth is played by the snapshot measurement made at a random phase of the spike light curve. Stated another way, the snapshot observation mode must necessarily result in a power-law distribution of the observed amplitudes. The fact, that such a power-law distribution is actually observed, can, thus, be interpreted as a successful consistency test of our spike decomposition algorithm.

Although some characteristics of the FST instrument (relatively narrow $1-1.5$~GHz frequency range and the low duty-cycle, snap-shot sampling) limit our use of the spike statistics for probing the corona, its other unique characteristic---directly sampling the time-domain data---has allowed us to investigate the performance of our statistically-based fitting algorithm for a range of realizations of the same spike data. The algorithm is found to be robust and should be directly applicable to data taken with other FFT-based radio instruments such as the Jansky Very Large Array and the Expanded Owens Valley Solar Array.  In addition, the theory and fitting algorithm built based on it have a much broader applicability range than fitting of solar radio spikes, as it applies to any spectrum obtained via FFT of time-domain signals.

\begin{appendices}

\section{Moment-based Approximation for the PDF of the Goodness of Fit Estimator}
\label{APP_CHI2PDF}
Although finding an exact closed form for the PDF of the random variable $\chi_N^2$ may be a difficult, if not impossible, mathematical task, the less challenging task of finding a approximation sufficient for practical applications may be straightforwardly accomplished by evaluating the expectation $E(\chi_N^2)=1$ and the higher statistical moments of the goodness of fit parameter, i.e. $E\left(\chi_N^{2k}\right)$.

For this purpose, we write
\begin{equation}
\label{tdchisqr}
\chi_N^2=\frac{1}{N}\sum_{j=1}^N\xi^2_j,
\end{equation}
where
\begin{equation}
\xi\equiv\sqrt{M}(\rho_j-1)
\end{equation}
is a random variable that, being linearly related to $\rho_j$, follows a re-scaled and translated $\mathscr{G}$ PDF, i.e.
\begin{equation}
\label{tdpdf}
\mathscr{G^*}(\xi)=\frac{\left(\sqrt{M}\right)^M \left(\xi+\sqrt{M} \right)^{M-1}}{\Gamma(M)}e^{-\sqrt{M}\left(\xi+\sqrt{M} \right)}.
\end{equation}

From this perspective, $\chi_N^2$ represents the sample mean of the squared random variable $\xi^2$, which allows us to use the PDF of $\xi$ for computing the expected variance of $\chi_N^2$ from a simple formula that relates it to the expectations of $\xi^4$ and $\xi^2$ \citep{RFI2},
\begin{equation}
\label{varchisr}
\sigma_{\chi_N^2}^2\equiv\frac{1}{N}\left[E(\xi^4)-E(\xi^2)^2\right]=\frac{2}{N}\left(1+\frac{3}{M}\right).
\end{equation}
This shows that, for a fixed $N$, the variance of the goodness of fit estimator asymptotically decreases as the accumulation length $M$ increases, from a maximum value of $8/N$ toward $2/N$, which is the variance of a standard $\chi_N^2$ PDF.

Having determined the mean and variance of the random variable equation~(\ref{chi2Inorm}) and, implicitly, of the goodness of fit estimator defined in equation~(\ref{redchisqr}), one may define an approximation for the yet unknown PDF by choosing a functional form that matches these first two standard moments.

A natural choice for such approximation is equation~(\ref{chisqPDFapprox}), which in the limit of large accumulation length $M$ reduces to a classic chi-squared distribution normalized by its degrees of freedom (equation~\ref{normchi2pdf}).

Within the limits of this approximation, the expected skewness, $\alpha_3$ and kurtosis, $\beta_2$, of the $\chi_\nu^2$ goodness of fit estimator are
\begin{eqnarray}
\alpha_3&\approx&\frac{2\sqrt{2}}{\sqrt{N}}\sqrt{1+\frac{3}{M}}\\\nonumber
\beta_2&\approx&3+\frac{12}{N}\left(1+\frac{3}{M}\right),
\end{eqnarray}
and the probability to observe a given $\chi_\nu^2$ or larger is given by Eq.~(\ref{pvalue})

\section{An Unbiased Estimator for the Variance of an Accumulated FFT Spectrum}
\label{sumvar}
If $\widehat{\sigma_j^2}$ is an unbiased estimator for the variance of the $\gdf(S_j;M,s_j/M)$ parent distribution of an accumulation of M raw FFT spectra, $S_j=\sum_{i=1}^M y_j$, then
\begin{equation}
E[\widehat{\sigma_j^2}]=s_j^2/M=M\mu_j^2,
\end{equation}
where $\mu_j$ and $\mu_j^2$ represent the mean and, respectively, the variance of the parent distribution $\gdf(S_j;1,\mu_j)$ corresponding to a raw FFT spectrum.

If $\widehat{\mu_j^2}$ represents the sample-based unbiased estimator of the variance of the $\gdf(S_j;1,\mu_j)$ distribution, i.e. $E[\widehat{\mu_j^2}]=\mu_j^2$. , immediately follows that
\begin{equation}
\label{sum2raw}
\widehat{\sigma_j^2}=M \widehat{\mu_j^2}.
\end{equation}
Hence, using the generally valid expression for the sample-based unbiased variance estimator \citep{Kendall}, $\widehat{\mu_j^2}$ may be written as
\begin{equation}
\label{est_sigma}
\widehat{\mu_j^2}=\frac{MS_2^{(j)}-S_j^2}{M(M-1)},
\end{equation}
where the $S_2^{(j)}$ represents the accumulated square of the power, $S_2^{(j)}=\sum_{i=1}^M y_j^2$, a quantity that is not routinely available from a standard spectrum analyzer, but due to its theoretically proven practical benefits \citep{RFI1, RFI2, RFI4}, it has recently become a standard output for a new generation of spectral instruments \citep{KSRBL, RFI3, Deller}.

However, even in the absence of directly measured $S_2^{(j)}$, by using an intermediate result of \citet{RFI2}, who showed that the statistical expectation of the accumulation $S_2^{(j)}$ is related to the expectation of the accumulation $S_j$,
\begin{equation}
\label{SK}
E[S_2^{(j)}]=\frac{2[E(S_j^2)]}{M+1},
\end{equation}
one may express the unbiased variance estimator $\widehat{\mu_j^2}$ solely in terms of the known accumulation $S_j$.

Indeed, by taking the expectation of the unbiased variance estimator given by Equation \ref{est_sigma}, one gets
\begin{eqnarray}
\mu_j^2=&&E\left[\widehat{\mu_j^2}\right]=
\frac{ME[S_2^{(j)}]-[E(S_j^2)]}{M(M-1)}=\\\nonumber
&&\frac{E(S_j^2)}{{M(M+1)}}=E\left[\frac{S_j^2}{{M(M+1)}}\right],
\end{eqnarray}
which proves that
\begin{equation}
\label{est_sigma2}
\widehat{\mu_j^2}=\frac{S_j^2}{{M(M+1)}}
\end{equation}
is an unbiased sample--based estimator for the variance of the $\gdf(S_j;1,\mu_j)$ distribution.

Consequently, from equation~(\ref{sum2raw}) it immediately follows that
\begin{equation}
\widehat{\sigma_j^2}=\frac{S_j^2}{{(M+1)}}
\end{equation}
is an unbiased sample--based estimator for the variance of the $\gdf(S_j;M,s_j/M)$ distribution, which describes the statistical properties of an accumulation of M FFT raw spectra.

\section{A Comparison between the Least-Squares and Maximum Likelihood Solutions}
\label{comparison}

The vanishing conditions of the first order partial derivatives of the spectral log-likelihood function with respect to its model parameters provide a system of $\nu$ equations with $\nu$ unknowns $p_k$,

\begin{equation}
\label{d1}
\frac{\partial\lambda(p_1,p_2...,p_\nu)}{\partial{p_k}}\equiv2M\sum_{j=1}^N\left(1-\frac{S_j}{\widehat{s_j}}\right)\frac{\partial\ln(\widehat{s_j})}{\partial{p_k}}=0.
\end{equation}
The formal uncertainties of the solution parameters may be expressed in terms of the second order partial derivatives of the log-likelihood function \citep{Bev} as

\begin{eqnarray}
\label{d2}
&&\sigma_{p_k}^2=2\left[\frac{\partial^2\lambda}{\partial{p_k^2}}\right]^{-1}=\\\nonumber
&&\frac{1}{M}\Big\{\sum_{j=1}^N\Big[\left(1-\frac{S_j}{\widehat{s_j}}\right)\frac{1}{\widehat{s_j}}\frac{\partial^2\widehat{s_j}}{\partial{p_k^2}}\\\nonumber
&&-\left(1-2\frac{S_j}{\widehat{s_j}}\right)\left(\frac{\partial\ln(\widehat{s_j})}{\partial{p_k}}\right)^2\Big]\Big\}^{-1}.
\end{eqnarray}

Since the system given by equation~(\ref{d1}) can be solved analytically only in some simple cases \citep{Bev}, the minimization of the spectral log-likelihood function generally requires a numerical method \citep[][and references therein]{xray}.

At this point, one may ask whether the functional form $\chi^{2(II)}$ defined in equation~(\ref{chisqr2}) may provide an alternative choice for a goodness-of-fit estimator. Indeed, using the probability density function $\gdf\left(\rho_j,M,\frac{1}{M}\right)$, one may compute the expectation
\begin{eqnarray}
\label{chisqII_expectation}
E\left[\chi^{2(II)}\right]&=&\frac{(M+1)(M+2)}{( M-1) (M-2)}N
\end{eqnarray}
and define the alternative statistical estimator
\begin{eqnarray}
\label{chi2IInorm}
&&\chi^{2(II)}_N\equiv\frac{( M-1) (M-1)}{(M+2)}\frac{1}{N}\sum_{j=1}^N\left(1-\frac{1}{\widehat{\rho_j}}\right)^2,
\end{eqnarray}
which has unity expectation. However, when compared with $\chi^{2(I)}_N$, its more cumbersome mathematical dependence on $\widehat{\rho_j}$ makes $\chi^{2(II)}_N$ a less suitable choice for a convenient goodness-of-fit estimator.

\begin{figure}
\epsscale{1}\plotone{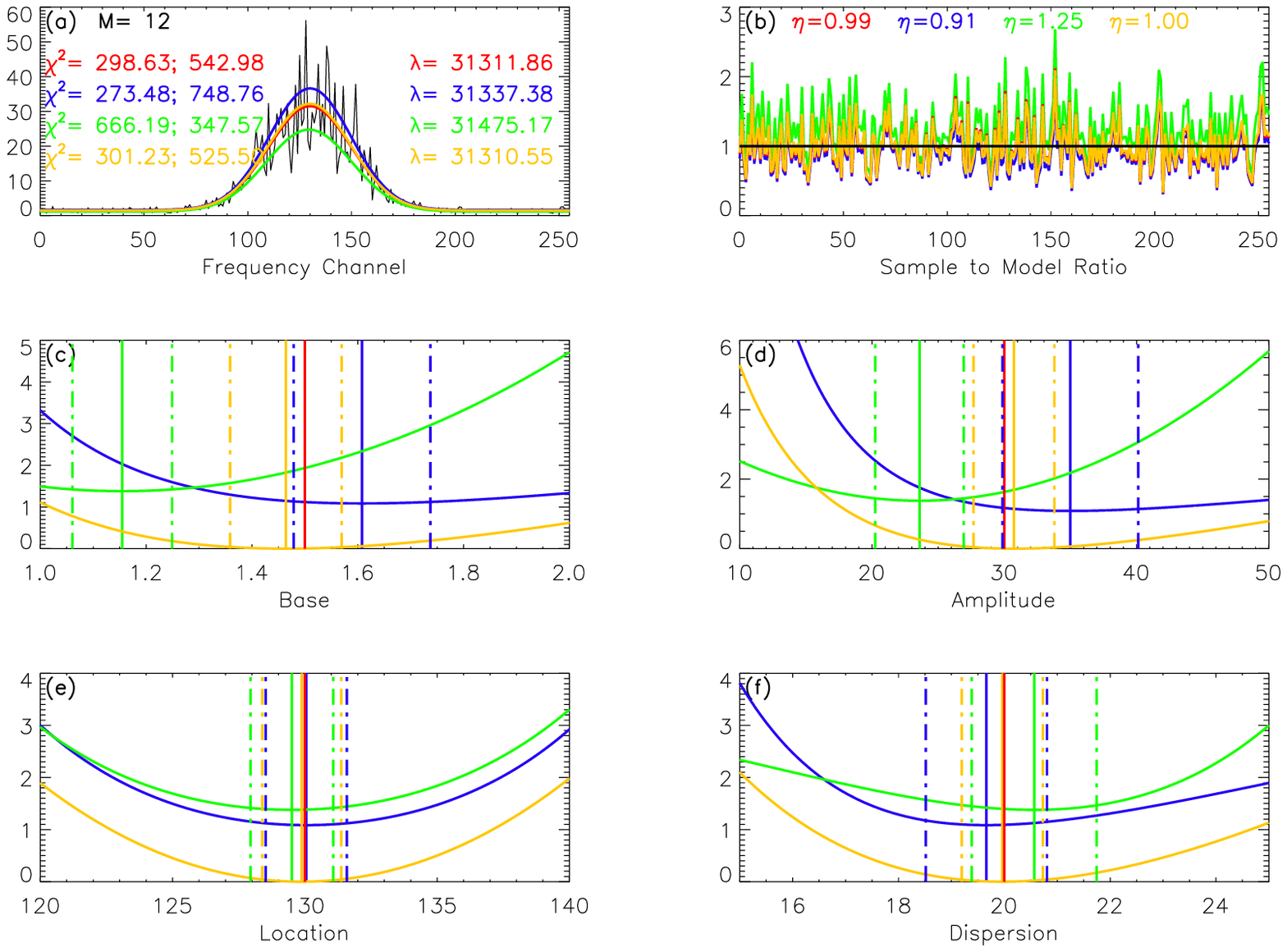}
\caption{\label{one} Fit results in the case of a simulated $M=12$ FFT spectrum containing a single Gaussian spectral peak superimposed on a flat background. The color-coded lines represent: red--true signal; blue--$\chi^2$ minimization according to Equation \ref{chisqr1}; green--$\chi^2$ minimization according to Equation \ref{chisqr2}; yellow--$\lambda$ minimization according to Equation \ref{ll}. The solutions, and their corresponding SMR values are shown on panels (a) and (b) respectively.  The associated $\chi_\nu^2$ (panel a), $\lambda$ (panel a), and $\eta$ (panel b) parameters are displayed in corresponding colors. The shapes of the minimized functions in the vicinity of their minima obtained by keeping three out of the four parameters fixed at their estimated values, while the remaining one has been slightly varied are shown in panels c--f. The $\chi^2$ curves have been normalized by their corresponding degrees of freedom, while their corresponding minimum values have been subtracted from the $\lambda$ curves. The true and estimated values of the spectral peak parameters are indicated in solid colors, while the $3\sigma$ standard ranges of uncertainty, computed according to the equations~(\ref{d2}) and (\ref{chi2sigma}), are indicated by color-coded dashed lines.}
\end{figure}

To check the validity of our expectation that the least-squares minimization may lead to a point in the parameter space that is located in the vicinity of the true maximum likelihood solution, we compare in Figure~\ref{one} the results obtained by minimizing the negative log--likelihood function (equation~\ref{ll}) and the two alternative $\chi^2$ functions (equations~\ref{chisqr1} and \ref{chisqr2}) in the case of a simulated $M=12$ FFT spectrum containing a single Gaussian spectral peak,
\begin{equation}
s=\xi+\alpha e^{-\frac{1}{2}\left(\frac{x-\beta}{\gamma}\right)^2},
\end{equation}
where $\alpha$, $\beta$, and $\gamma$, i.e. the amplitude, location, and dispersion, respectively, of the Gaussian peak, and $\xi$ represents a flat spectral background. The true signal, and the peaks estimated by these three alterative methods are shown in panel (a).

Based on Figure~\ref{one}b we may conclude, by visual inspection, that the log--likelihood minimization provides a more accurate representation of the hidden true signal than the $\chi^{2(I)}$ or $\chi^{2(II)}$ estimations, despite the fact that they correspond to smaller least--squared deviations. The fact that the minimization of $\chi^{2(I)}$ results in an overestimation of the true amplitude $\alpha$, while the minimization of $\chi^{2(II)}$  underestimates it, may be understood as a direct consequence of their reciprocal mathematical definitions, which result in different weighting of the data points. At the same time, the minimized $\lambda$ parameter and the non-minimized $\chi^{2(I)}$ and $\chi^{2(II)}$ values corresponding to the negative log--likelihood estimation are closer to the true values produced by the random realization used for this test.

Figure~\ref{one}b displays the local SMR deviations corresponding to the noise--contaminated true signal and the three alternative estimations, as well as the average SMR deviations, $\widehat{\eta}=(1/N)\sum_{j=1}^N\widehat{\rho_j}$. While the lower and, respectively, higher than unity averaged SMR values resulting from the two $\chi^2$  minimizations are related to the same weighting bias discussed above, the perfect $\widehat{\eta}=1.00$ from the log--likelihood minimization is consistent with the known fact that the log-likelihood approach may not necessarily provide the most accurate description of an individual random realization, but does provide the statistically most favorable set of parameters that could have produced the observed samples.

To provide the means for a more quantitative assessment of this test, we display in the other four panels of the Figures \ref{one} the shapes of the minimized functions in the vicinity of their minima obtained by keeping three out of the four parameters fixed at their estimated values, while the remaining one has been slightly varied. To facilitate a direct comparison, the $\chi^2$ curves have been normalized by their corresponding degrees of freedom, while their corresponding minimum values have been subtracted from the $\lambda$ curves. The true and estimated values of the spectral peak parameters are indicated in solid colors, while the $3\sigma$ standard ranges of uncertainty, computed according to the equations~(\ref{d2}) and (\ref{chi2sigma}), are indicated by color-coded dashed lines.

\section{ Standard Moments of the Mean of an SMR Compact Region }
\label{appendix_moments}

One may compute the characteristic functions corresponding to the probability distribution functions describing the means $\eta_a$ and $\eta_b$ by taking the $n^{th}$ power of the Fourier transforms of the conditional PDFs defined by equation~(\ref{cnormpdf}), followed by the standard change of variable $t\rightarrow t/n$ in their arguments \citep{Kendall}. This procedure leads to
  \begin{eqnarray}
  \label{cf}
  \Phi(t)\Big|_{\{\rho_j>1\}}&=&\left[\frac{M^M E(1-M,M-\frac{i t}{n})}{\gamma(M,M)}\right]^n\\\nonumber
  \Phi(t)\Big|_{\{\rho_j<1\}}&=&\left[\frac{M^M (M-\frac{i t}{n})^{-M}[\Gamma(M)-\gamma(M,M-\frac{i t}{n})]}{\gamma(M,M)}\right]^n,
 \end{eqnarray}
where
\begin{equation}
E(a,z)=\int_1^{\infty}e^{-zt}t^{-a}dt,
\end{equation}
is the standard exponential integral function.

Although closed forms for the inverse Fourier transforms of equations~(\ref{cf}) have not been found, which would have directly provided the PDF we are interested in, by following a standard procedure \citep{Kendall}, one may use these characteristic functions to compute the infinite sets of statistical moments associated with the random variables under investigation.

Using the characteristic functions given by equation~(\ref{cf}), one may straightforwardly compute the first four central moments of the random variables $\eta_a$ and $\eta_b$ (equation~\ref{sab}), which are given by equations~(\ref{cma}) and (\ref{cmb}), respectively.
\begin{eqnarray}
\label{cma}
\mu&=&\frac{\Gamma(M+1,M)}{M\gamma(M,M)}\\\nonumber
\mu_2&=&\frac{\gamma(M,M)\Gamma(M+2,M)-\Gamma(M+1,M)^2}{nM^2\gamma(M,M)^2}\\\nonumber
\mu_3&=&\frac{1}{n^2M^3\gamma(M,M)^3}\Big[2\Gamma(M+1,M)^3\\\nonumber
      &&-3\gamma(M,M)\Gamma(M+1,M)\Gamma(M+2,M)\\\nonumber
      &&+\gamma(M,M)^2\Gamma(3+M,M)\Big]\\\nonumber
\mu_4&=&\frac{1}{n^3M^4\gamma(M,M)^4}\Big[3(n-2)\Gamma(M+1,M)^4\\\nonumber
      &&-6(n-2)\gamma(M,M)\Gamma(M+1,M)^2\Gamma(M+2,M)\\\nonumber
      &&-4\gamma(M,M)^2\Gamma(M+1,M)\Gamma(M+3,M)\\\nonumber
      &&+3(n-1)\gamma(M,M)^2\Gamma(M+2,M)^2\\\nonumber
      &&+\gamma(M,M)^3\Gamma(M+4,M)\Big]
\end{eqnarray}

\begin{eqnarray}
\label{cmb}
\mu&=&1-\frac{e^{-M}M^{M-1}}{\Gamma(M)-\gamma(M,M)}\\\nonumber
\mu_2&=&\frac{e^{-2M}}{n M^2[\Gamma(M)-\gamma(M,M)]^2}\Big\{e^{2M}M\Gamma(M)^2\\\nonumber
       &&+e^MM^M\gamma(M,M)+e^{2M}M\gamma(M,M)^2\\\nonumber
       &&-{e^M}\Gamma(M)[M^M+2e^{M}M\gamma(M,M)]-M^{2M}\Big\}\\\nonumber
\mu_3&=&\frac{e^{-3M}}{n^2M^3[\Gamma(M]-\gamma(M,M)]^3}\\\nonumber
      &&\times\Big\{2e^{3M}M\Gamma(M)^3+3e^MM^{2M}\gamma(M,M)\\\nonumber
      &&+e^{2M}(M-2)M^M\gamma(M,M)^2\\\nonumber
      &&-2e^{3M}M\gamma(M,M)^3-2M^{3M}\\\nonumber
      &&-e^{2M}\Gamma(M)^2\Big[6e^MM\gamma(M,M)-(M-2)M^M\Big]\\\nonumber
      &&+e^M\Gamma(M)\Big[6e^{2M}M\gamma(M,M)^2\\\nonumber
      &&-2e^M(M-2)M^M\gamma(M,M)-3M^{2M}\Big]\Big\}\\\nonumber
\mu_4&=&\frac{e^{-4M}}{n^3M^4[\Gamma(M]-\gamma(M,M)]^4}\\\nonumber
       &&\times\Big\{3(n-2)M^{4M}+3e^4MM(M n+2)\Gamma(M)^4\\\nonumber
       &&-6(n-2)e^MM^{3M}\gamma(M,M)\\\nonumber
       &&-e^{2M}M^{2M}[M(6n-4)-3n+11]\gamma(M,M)^2\\\nonumber
       &&+e^{3M}M^M[(6n-5)M+6]\gamma(M,M)^3\\\nonumber
       &&+3e^{4M}M(M n+2)\gamma(M,M)^4\\\nonumber
       &&-e^{3M}\Gamma(M)^3\Big[M^M[(6n-5)M+6]\\\nonumber
       &&+12e^M(M n+2)\gamma(M,M)\Big]\\\nonumber
       &&+e^{2M}\Gamma(M)^2\Big[M^{2M}[(6n-4)M+3n-11]\\\nonumber
       &&+3e^MM^M[(6n-5)M+6]\gamma(M,M)\\\nonumber
       &&+18e^{2M}M(M n+2)\gamma(M,M)^2\Big]\\\nonumber
       &&-e^M\Gamma(M)\Big[6(2-n)M^{3M}-2e^{M}M^{2M}\\\nonumber
       &&\times[(6n-4)M-3n+11)\gamma(M,M)\\\nonumber
       &&3e^{2M}M^M[(6n-5)M+6]\Gamma[M,M]^2\\\nonumber
       &&12e^{3M}M(M n+2)\gamma(M,M)^3\Big]\Big\}
\end{eqnarray}

\section{Pearson Type I Approximations vs. Monte Carlo Simulations}
\label{appendix_P1vMC}

One should always take in consideration that the Pearson Type I are only approximations to the true distributions, therefore their performance and limitations should be always tested on a case-by-case basis.

To facilitate such critical assessment, we present in Table \ref{betaparmstable} the inferred Pearson Type I parameters for the particular case $M=48$ used in the Monte Carlo experiment described in \S~\ref{S_Sim}. The parameters shown for each group are the skewness, $\alpha_3$, the allowed range $\{a,b\}$, and the $0.13499\%$ probability thresholds $t_a$ an $t_b$, as defined in \S~\ref{P1}.

In addition, for each group of distributions, the \emph{tunneling} probabilities for $\eta_b$ and $\eta_a$ to be greater and, respectively, lesser the unity are given. The fact that these probabilities are non-zero is the direct consequence of the fact that both $\eta_b$  and $\eta_a$ Pearson I approximations have ranges that actually cross above and respectively below unity, although it is mathematically impossible to observe in practice means above or below unity while averaging quantities that are all below and respectively above these borders.

However, except for the special case $n=1$, these pseudo tunneling probabilities are below unity and their negligibility increases as fast as one order of magnitude as $n$ increases by 1, despite the fact that neither the distribution limits nor the relative tunneling depths changes noticeably with $n$ as $M$ is kept fixed. These results suggests that the accuracy of these approximations may increase with $n$.

The special case $n=1$ may be easily explained by recalling the fact that for $n=1$ we already found the true distributions, which are the truncated $Gamma$ distributions given by equation~(\ref{cnormpdf}), which we have used as basis for our derivation. The $n=1$ distributions, despite having the firsts standard moments satisfying the Pearson Type I region criterion, as clearly illustrated in Figure~\ref{pearsonregions}, do not not satisfy the basic Pearson family criterion of having a continuous derivative. Therefore, the available true distributions must be used for the case $n=1$ and, since in all other cases the true distributions are not known, we compare them with the outcome of the Monte Carlo simulations in order to validate the Pearson Type I approximations derived for $n>1$.
The result of this comparison, which we present in this Appendix, allows us to conclude that, at least from a practical point of view, the true tail probabilities of the SMR distributions are accurately estimated by the Pearson Type I Approximations in all cases including the special case $n=1$.

\begin{figure}
\epsscale{0.8}\plotone{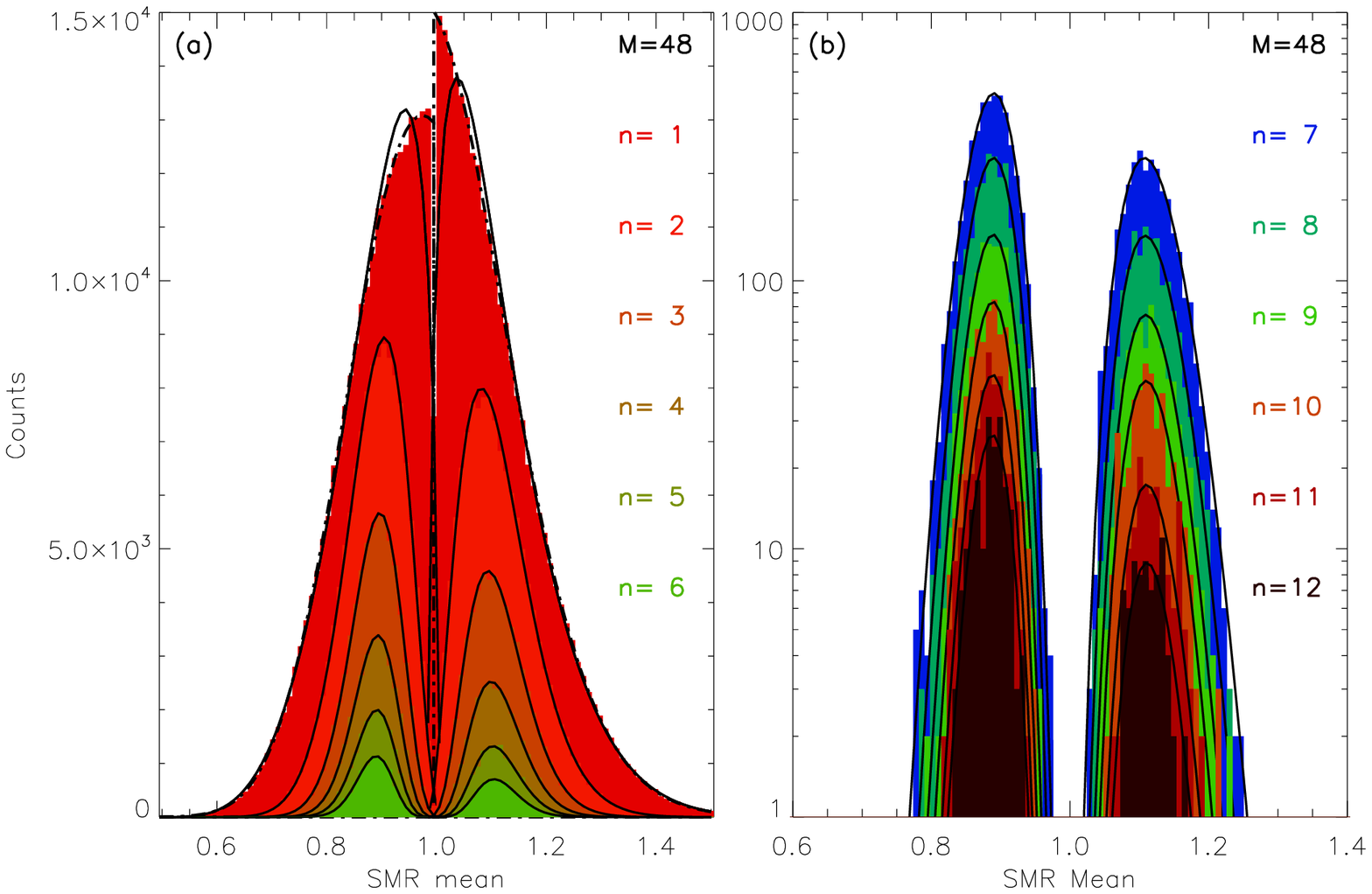}
\caption{\label{betadistributions} The observed simulated distributions of compact SMR regions of different sizes $n$ and fixed accumulation length $M$. The good agreement with the Pearson type I approximations is evident,(note the logarithmic scale on panel $(b)$), for all cases except $n=1$ which is exactly described by the truncated $Gamma$ distributions (Eqn. \ref{cnormpdf}) indicated on panel $a$ by the dashed-dotted lines.}
\end{figure}

Figure~\ref{betadistributions}a shows the randomly generated distributions of the mean SMR $s_{b}$ and $\eta_a$ for $n=\overline{1,6}$, while Figure~\ref{betadistributions}b shows the distributions for $n=\overline{7,12}$. For comparison, their corresponding Pearson Type I PDF approximations, characterized by the parameters shown in Table~\ref{betaparmstable} are drawn with solid lines. A very good agreement between simulations and theoretical approximations is evident except for the special case $n=1$, for which the corresponding true theoretical distributions are shown as dashed-dotted lines. Although, for reasons already discussed in the precedent section, the differences between the $n=1$ approximations and the true distributions are significant especially in the vicinity of their peaks, Figure \ref{betadistributions}a suggests that, even in the case $n=1$, the Pearson Type I curves may still offer good approximations for the true tail probabilities. To support this assertion, we present in Table \ref{comptable} a comparison between some key parameters of the approximative and exact distributions for $M=48$ and $n=1$. Table \ref{comptable} reveals that the standard $0.13499\%$  PFA thresholds, $t_a$ and $t_b$, computed using the $Beta$ approximations are practically identical with the thresholds computed using the exact distributions. Moreover, the true probabilities of observing SMR regions beyond the limited ranges of the $Beta$ approximations are also practically negligible.

\begin{deluxetable}{lrrrrrrrrrrr}
\tablecolumns{6}
\tablewidth{0pc}
\tablecaption{\label{comptable}Exact PDFs vs. Pearson Type I Approximations for $M=48$ and $n=1$}
\tablehead{
\colhead{Type}&\colhead{$p(s<a)\%$}&\colhead{a}&\colhead{$t_b(0.13\%)$}&\colhead{b}&\colhead{$p(s>1)\%$}\\
\cline{1-6}\\\multicolumn{6}{c}{Compact Regions Below Unity}
}
\startdata
Pearson Type I&    0.000&0.4852&0.6079&1.0106&     2.14\\
Truncated Gamma& 0.000001&\nodata&0.6018&1.0000&    0.000\\
\cutinhead{Compact Regions Above Unity}
\colhead{Type}&\colhead{$p(s<a)\%$}&\colhead{a}&\colhead{$t_a(0.13\%)$}&\colhead{b}&\colhead{$p(s>1)\%$}\\
\cline{1-6}\\
Pearson Type I&     2.44&0.9895&1.5276&2.0579&    0.000\\
Truncated Gamma&    0.000&1.0000&1.5299&\nodata& 0.000943\\
\enddata
\end{deluxetable}

From a practical point of view, to address the fitting problem, we are interested only in the tail probabilities. Therefore, we may assert that even for the case $n=1$ it would be safe to assume that any outlier located outside the range of the Pearson Type I distributions would be due to the model function rather than the result of some expected statistical fluctuations. This result may greatly simplify and speed up the fitting algorithm since, instead of paying the computational cost of having to evaluate an incomplete $Beta$ function for checking the statistical significance of each observed systematic deviation above or below unity, and compare it with an arbitrary chosen non zero probability threshold, one may chose instead to compute only the limited ranges of the Pearson Type I approximations, such as those listed by Table \ref{betaparmstable} , and use them to reject any systematic region that is situated beyond these limits.

\begin{figure}
\epsscale{0.8}\plotone{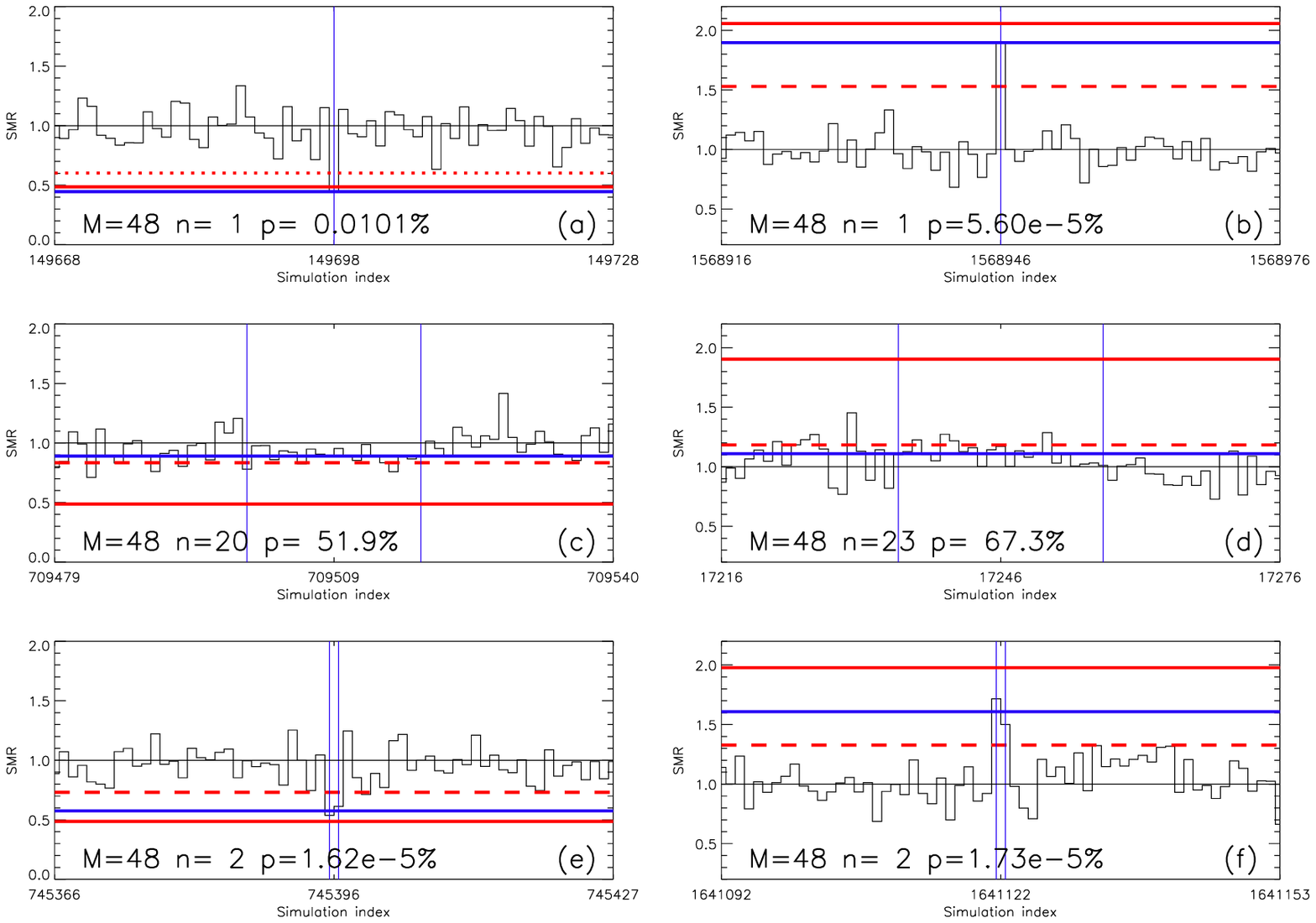}
\caption{\label{simextreemes} Simulation highlights: a) the region containing the largest SMR deviation below unity; b) the region containing the largest SMR deviation above unity; c) the largest region displaying a systematic deviation below unity; d) the largest region displayed a systematic deviation above unity; e) the least probable region below unity; f) the least probable region above unity. The ranges of the selected regions are indicated by vertical blue solid lines and their corresponding mean deviations are shown by horizontal blue solid lines on each panel. The red dashed horizontal lines indicate the standard $0.13499\%$ thresholds corresponding to an individual deviation, and the red solid lines indicate the maximum range of the deviations allowed in each case by the Pearson Type I approximations. On each panel, the region length $n$, the size probability, $p(n)$, and the SMR mean probability $p(s)$ to be observed amongst its own size class are also indicated.}
\end{figure}

To conclude this section, we present in Figure \ref{simextreemes} two data sequences corresponding to the simulated data regions, that contain the largest SMR deviations below (0.554--panel a) and above (0.897--panel b)unity, the largest systematic SMR regions below ($n=20$--panel c) and above ($n=23$--panel d) unity, as well as the least probable SMR regions randomly generated below ($2.11\times10^{-4}\;\%$--panel e) and above ($2.08\times10^{-4}\;\%$--panel f) unity.

The ranges of the selected regions are indicated by blue vertical lines and their mean deviations are shown as horizontal blue solid lines on each panel. The red dashed horizontal lines indicate the corresponding standard $0.13499\%$ thresholds and the red solid lines indicate, in each case, the maximum deviations from unity allowed by the Pearson Type I approximations. On each panel, the region length $n$, the size probability, $p(n)$, and the SMR mean probability $p(s)$ to be observed amongst its own size class are also indicated.

Note that, in a real situation, both $n=1$ SMR regions shown in panels $(a)$ and $b$ would fail the standard $0.13499\%$ PFA test. However, unlike the panel $a$ region, the panel $b$ region would also survive the $0\%$ PFA test, being still inside the limited boundaries of its Pearson Type I approximation, despite its comparatively lower probability to be observed amongst its own class. At the same time, despite being very unlikely to be randomly generated, as resulted from their size distributions (Eqn. \ref{nab}), i.e. $9.74\times10^{-5}\;\%$ and $2.52\times10^{-6}\;\%$, respectively, the regions shown in panels $(c)$ and $(d)$ would survive both standard and  boundary tests due to their high probabilities to be observed amongst their own class, i.e. $51.9\%$ and $67.3\%$, respectively, as resulted from Eqn. \ref{betaprob}.

Hence, combining the probabilities provided by Eqn. \ref{nab} and Eqn. \ref{betaprob}, one may compute a relative ranking in the increasing order of their absolute probabilities to being randomly generated, i.e. $(d): 1.69\times10^{-4}\;\%,\; (f): 2.08\times10^{-4}\;\%,\; (e): 2.10\times10^{-4}\;\%,\; (a): 2.52\times10^{-3}\;\%$, and $ (c):\; 5.06\times10^{-3}\;\%$. This reveals that the largest $(n=23)$ SMR region shown in panel $(d)$ is indeed the least likely region to be randomly generated, while the $n=20$ region shown in panel $c$ is actually the one that is most likely to be randomly generated, despite being much larger than the $n=2$ regions.

We consider as a significant outcome of our simulation the fact that out of a total of $498,611$ SMR regions having $n\ge2$, none has been found to cross the limited boundaries of its corresponding Pearson Type I PDF Approximation. Moreover, out of a total of $500,343$ single element SMR regions, only $2$ below unity and none above unity have been found to cross the boundaries of their corresponding Pearson Type I PDF Approximations.

This results indicates that, at least from a practical point of view, the true tail probabilities of the SMR distributions are accurately estimated by the Pearson Type I Approximations in all cases including the special case $n=1$.
\end{appendices}

\acknowledgments
This work was supported in part by NSF grants
AGS-1250374 and NASA grants  NNX11AB49G and NNX13AE41G to New
Jersey Institute of Technology. This work also benefited from workshop support from the
International Space Science Institute (ISSI).

\begin{deluxetable}{rrrrrrrrrrrr}
\tablecolumns{12}
\tablewidth{0pc}
\tablecaption{\label{betaparmstable}Parameters of the Pearson Type I Approximations for $M=48$ and $n=\overline{1,25}$}
\tablehead{
&\multicolumn{5}{c}{Compact Regions Below Unity}  &&\multicolumn{5}{c}{Compact Regions Above Unity} \\ \cline{2-6} \cline{8-12} \\
\colhead{n}&\colhead{$\alpha_3$}&\colhead{a}&\colhead{$t_b$}&\colhead{b}&\colhead{$p(s>1)$}& &\colhead{$p(s<1)$}&\colhead{a}&\colhead{$t_a$}&\colhead{b}&\colhead{$\alpha_3$}\\
&&&\colhead{$@0.13\%$}&&\colhead{\%}&&\colhead{\%}&&\colhead{$@0.13\%$}
}
\startdata
       1&-0.81&0.4852&0.6079&1.0106&     2.14&&     2.44&0.9895&1.5276&2.0579& 1.17\\
       2&-0.57&0.4863&0.6927&1.0158& 2.e-001&& 2.e-001&0.9876&1.3854&1.9773& 0.83\\
       3&-0.47&0.4866&0.7313&1.0175& 1.e-002&& 1.e-002&0.9869&1.3280&1.9505& 0.68\\
       4&-0.40&0.4868&0.7541&1.0183& 1.e-003&& 1.e-003&0.9865&1.2955&1.9371& 0.59\\
       5&-0.36&0.4869&0.7695&1.0188& 1.e-004&& 9.e-005&0.9863&1.2740&1.9290& 0.52\\
       6&-0.33&0.4869&0.7808&1.0192& 1.e-005&& 8.e-006&0.9861&1.2586&1.9237& 0.48\\
       7&-0.31&0.4870&0.7894&1.0194& 1.e-006&& 7.e-007&0.9860&1.2469&1.9199& 0.44\\
       8&-0.29&0.4870&0.7964&1.0196& 1.e-007&& 6.e-008&0.9859&1.2376&1.9170& 0.41\\
       9&-0.27&0.4870&0.8021&1.0197& 9.e-009&& 6.e-009&0.9859&1.2299&1.9148& 0.39\\
      10&-0.26&0.4871&0.8069&1.0199& 9.e-010&& 5.e-010&0.9858&1.2236&1.9130& 0.37\\
      11&-0.24&0.4871&0.8110&1.0199& 9.e-011&& 5.e-011&0.9858&1.2182&1.9115& 0.35\\
      12&-0.23&0.4871&0.8145&1.0200& 9.e-012&& 4.e-012&0.9857&1.2135&1.9103& 0.34\\
      13&-0.22&0.4871&0.8177&1.0201& 8.e-013&& 4.e-013&0.9857&1.2094&1.9093& 0.32\\
      14&-0.22&0.4871&0.8205&1.0201& 8.e-014&& 3.e-014&0.9857&1.2058&1.9084& 0.31\\
      15&-0.21&0.4871&0.8229&1.0202& 8.e-015&& 3.e-015&0.9857&1.2025&1.9076& 0.30\\
      16&-0.20&0.4871&0.8252&1.0202& 8.e-016&& 3.e-016&0.9856&1.1996&1.9070& 0.29\\
      17&-0.20&0.4871&0.8272&1.0203& 8.e-017&& 3.e-017&0.9856&1.1970&1.9064& 0.28\\
      18&-0.19&0.4872&0.8291&1.0203& 8.e-018&& 2.e-018&0.9856&1.1946&1.9059& 0.28\\
      19&-0.19&0.4872&0.8308&1.0203& 8.e-019&& 2.e-019&0.9856&1.1924&1.9054& 0.27\\
      20&-0.18&0.4872&0.8323&1.0204& 8.e-020&& 2.e-020&0.9856&1.1904&1.9050& 0.26\\
      21&-0.18&0.4872&0.8338&1.0204& 8.e-021&& 2.e-021&0.9856&1.1885&1.9046& 0.26\\
      22&-0.17&0.4872&0.8351&1.0204& 8.e-022&& 2.e-022&0.9856&1.1868&1.9043& 0.25\\
      23&-0.17&0.4872&0.8364&1.0204& 8.e-023&& 2.e-023&0.9855&1.1852&1.9039& 0.24\\
      24&-0.17&0.4872&0.8376&1.0204& 8.e-024&& 2.e-024&0.9855&1.1837&1.9036& 0.24\\
      25&-0.16&0.4872&0.8387&1.0205& 8.e-025&& 1.e-025&0.9855&1.1823&1.9034& 0.23 
\enddata
\end{deluxetable}


\end{document}